\documentclass{aa}
\usepackage{graphicx}

\def\ltsim{\raise 2pt \hbox {$<$} \kern-1.1em \lower 4pt \hbox {$\sim$}}
\def\ltapprox{\raise 2pt \hbox {$<$} \kern-1.1em \lower 5pt \hbox {$\approx
$}}
\def\gtsim{\raise 2pt \hbox {$>$} \kern-1.1em \lower 4pt \hbox {$\sim$}}
\def\gtapprox{\raise 2pt \hbox {$>$} \kern-1.1em \lower 5pt \hbox {$\approx
$}}

\begin{document}

\title{GMRT Radio Halo Survey in galaxy clusters at z = 0.2 -- 0.4.}
\subtitle{I.The REFLEX sub--sample} 
\author{T.~Venturi\inst{1} \and
S.~Giacintucci\inst{2,}\inst{1,}\inst{3} \and 
G.~Brunetti\inst{1}\and
R.~Cassano\inst{3,}\inst{1}\and
S.~Bardelli\inst{2} \and
D.~Dallacasa\inst{3,}\inst{1}\and 
G.~Setti\inst{3,}\inst{1}
}

\institute
{
INAF -- Istituto di Radioastronomia, via Gobetti 101, I-40129, Bologna, Italy 
\and
INAF -- Osservatorio Astronomico di Bologna,
via Ranzani 1, I--40127 Bologna, Italy
\and
Dipartimento di Astronomia, Universit\`a di Bologna,
via Ranzani 1, I--40126, Bologna, Italy
}
\date{}
%
%
\abstract
{}
{We present the first results of an ongoing project 
devoted to the search of giant radio halos in galaxy clusters
located in the redshift range z=0.2--0.4. One of the main goals of our
study is to measure the fraction of massive galaxy clusters in this
redshift interval hosting a radio halo, and to 
constrain the expectations of the particle re--acceleration model
for the origin of non--thermal radio emission in galaxy clusters.} 
{We selected 27 REFLEX clusters and here we present Giant Metrewave Radio 
Telescope (GMRT) observations at 610 MHz for 11 of them. 
The sensitivity (1$\sigma$) in our images is in the range 
35--100~$\mu$Jy beam$^{-1}$  for all clusters.}
{We found three new radio halos, doubling the number of halos known in the 
selected sample. In particular,
giant radio halos were found in A\,209 and RXCJ\,2003.5--2323, and one
halo (of smaller size) was found in RXCJ\,1314.4--2515. 
Candidate extended emission on smaller scale was found around the central 
galaxy in A\,3444 which deserves further investigation. 
Furthermore, a radio relic was found in A\,521, and two relics
were found in RXCJ\,1314.5--2515.
The remaining six clusters observed do not host extended emission of 
any kind.}
{}


\maketitle
\section{Introduction}\label{sec:intro}

Radio and X--ray observations of galaxy clusters prove 
that thermal and non--thermal plasma components coexist in the 
intracluster medium (ICM).
While X--ray observations reveal the presence of diffuse 
hot gas, the existence of extended cluster--scale radio 
sources in a number of galaxy clusters, well known as 
{\it radio halos} and {\it relics}, prove the presence of 
relativistic electrons and magnetic fields.
\\
Both radio halos and relics are low surface 
brightness sources with steep radio spectra, whose linear size 
can reach and exceed the Mpc scale. 
Radio halos are usually located at the centre of galaxy
clusters, show a fairly regular radio morphology, and 
lack an obvious optical counterpart.
A total of about 20 radio halos have been detected up to now
(Giovannini, Tordi \& Feretti \cite{giovannini99}; Giovannini \& Feretti 
\cite{gf02}; Kempner \& Sarazin \cite{kempner01}; 
Bacchi et al. \cite{bacchi03}). Relics are usually found at
the cluster periphery, their radio emission is highly polarized
(up to $\sim$ 30\%), and shows a variety of radio morphologies,
such as sheet, arc, toroids. At present
a total of $\sim$ 20 relics (including candidates) are known
(Kempner \& Sarazin \cite{kempner01}; Giovannini \& Feretti
\cite{gf04}).
\\
\\
Evidence in the optical and X--ray bands has been 
accumulated in favour of the hierarchical formation of galaxy clusters 
through merging processes (for a collection of reviews on this subject 
see Feretti, Gioia \& Giovannini 2002), and this 
has provided insightful pieces of information in our understanding of 
radio halos. It is not clear whether all clusters with signatures
of merging processes also possess a radio halo; on the other hand, 
all clusters hosting a radio halo show sub--structures in the X--ray emission,
and the most powerful radio halos are hosted in clusters 
which most strongly depart from virialization (Buote \cite{buote01}).
Giovannini et al. (1999) showed that in the redshift interval 0 -- 0.2
the detection rate of cluster radio halos increases with increasing X--ray 
luminosity, which suggests a connection with the gas temperature and cluster mass.
\\
\\
The very large extent of radio halos poses the question of their origin,
since the diffusion time the relativistic electrons need to cover the
observed Mpc size is 30 -- 100 times longer than their radiative lifetime.
Two main possibilities have been investigated so far: ``primary models'', in which 
particles are in--situ re--accelerated in the ICM, and ``secondary models'' 
in which the emitting electrons are secondary products of hadronic collisions
in the ICM (for reviews on these models see Blasi \cite{blasi04}; 
Brunetti \cite{brunetti03} and \cite{brunetti04}; Ensslin \cite{ensslin04}; 
Feretti \cite{feretti03}; Hwang \cite{hwang04}; Sarazin \cite{sarazin02}).
Cluster mergers are among the most energetic events in the Universe, with 
an energy release up to 10$^{64}$ erg, and a challenging question is if at 
least a fraction of such energy may be channelled into particle reacceleration
(e.g. Tribble \cite{tribble93}).
Indeed observational support (for a review see Feretti \cite{feretti03}) is now 
given to the particle re--acceleration model, which assumes that the 
radiating electrons are stochastically re--accelerated by turbulence in the 
ICM and that the bulk of this turbulence is injected during cluster mergers
(Brunetti et al. \cite{brunetti01}; Petrosian \cite{petrosian01};  
Fujita, Takizawa \& Sarazin \cite{fujita03};
Brunetti et al. \cite{brunetti04b}). 
\\
\\
Although the physics of particle re--acceleration by turbulence has been
investigated in some detail and the model expectations seem to reproduce
the observed radio features, only recently statistical calculations 
in the framework of the re--acceleration model have been carried out by
Cassano \& Brunetti (\cite{cassano05}, hereinafter CB05).
Making use of semi--analytical calculations they estimated the energy of turbulence
injected in galaxy clusters through cluster mergers, and derived the
expected occurrence of {\it giant}\footnote{Linear size  $\ge$ 1 Mpc as defined in CB05, with 
H$_0$=50 ~km~s$^{-1}$~Mpc$^{-1}$. 
This size corresponds to \gtsim 700 kpc with the cosmology assumed in 
this paper, i.e. $H_0 =70$~km~s$^{-1}$~Mpc$^{-1}$, $\Omega_m$=0.3 and 
$\Omega_{\Lambda}$=0.7. } radio halos as a function of the mass 
and dynamical status of the clusters in the framework of the merger--induced  
particle re--acceleration scenario.
\\
The most relevant result of those calculations is that the occurrence of giant
radio halos increases with the cluster mass. Furthermore, the expected fration
of clusters with giant radio halos at z$\le$ 0.2 can be reconciled with the 
observed one (Giovannini et al. \cite{giovannini99}) for viable values of the
model parameters.
\\
Cassano, Brunetti \& Setti (\cite{cassano04}, hereinafter CBS04) and 
Cassano, Brunetti \& Setti (\cite{cassano06}, hereinafter CBS06) showed
that the bulk of giant radio halos 
are expected in the redshift range $z\sim 0.2 \div 0.4$
as a result of two competing effects, i.e. the decrease of the fraction of clusters 
with halos in a given mass range and the increase of the volume of the Universe with 
increasing redshift. Given that inverse Compton losses increase with  the redshift, 
it is expected that powerful giant radio halos at $z>0.2$ are preferentially 
found in massive clusters ($M \sim 2-3 \times 10^{15}M_{\odot}$) 
undergoing merging events. In particular, it is expected that 
a fraction of 10 -- 35 \% of clusters in this redshift interval and mass range 
may host a giant radio halo.
\\
\\
With the aim to investigate the connection between cluster mergers and the
presence of cluster--type radio sources, in particular to derive
the fraction of massive galaxy clusters in the range 0.2 $<$ z $<$ 0.4
hosting a radio halo and constrain the predictions of the re--acceleration 
model in the same redshift interval, we undertook an observational
study using the Giant Metrewave Radio Telescope (GMRT, Pune, 
India) at 610 MHz. Our project will be presented here and in future papers,
and will be referred to as the GMRT Radio Halos Survey.
\\
Here we report the results on 11 galaxy clusters observed with the GMRT in 
January 2005. The paper is organised as follows: 
in Section \ref{sec:sample} we present the sample of galaxy
clusters; the radio observations are described in Section \ref{sec:obs}; 
the analysis of our results and a brief discussion are given in 
Section \ref{sec:results} and \ref{sec:discussion} respectively. 
\\
\\

\section{The cluster sample}\label{sec:sample}

%
%
\begin{table*}[t]
\label{tab:sample1}
\caption[]{Cluster sample from the REFLEX catalogue.}
\begin{center}
\begin{tabular}{rrccccrcc}
\hline\noalign{\smallskip}
REFLEX Name   & Alt. name & RA$_{J2000}$ &  DEC$_{J2000}$ & z &  L$_{\rm X}$(0.1--2.4 keV) 
& M$_{\rm V}$ & R$_{\rm V}$  \\ 
              &           &              &                &   & $10^{44}$ erg s$^{-1}$ 
& 10$^{15}$M$_{\odot}$ & Mpc  \\
\noalign{\smallskip}
\hline\noalign{\smallskip}
$^{\surd}$ RXCJ\,0003.1$-$0605 & A\,2697 &  00 03 11.8 &  $-$06 05 10 & 0.2320 &   6.876 
& 1.68 & 2.70    \\
$^{\star}$ RXCJ\,0014.3$-$3023 & A\,2744 &  00 14 18.8 &  $-$30 23 00 & 0.3066 &  12.916 
& 2.58 & 2.99    \\
$^{\surd}$ RXCJ\,0043.4$-$2037 & A\,2813 &  00 43 24.4 &  $-$20 37 17 & 0.2924 &   7.615 
& 1.80 & 2.67    \\
$^{\surd}$ RXCJ\,0105.5$-$2439 & A\,141 &  01 05 34.8 &  $-$24 39 17 & 0.2300 &   5.762 
& 1.50 & 2.60    \\ 
$^{\surd}$ RXCJ\,0118.1$-$2658 & A\,2895 &  01 18 11.1 &  $-$26 58 23 & 0.2275 &   5.559 
& 1.45 & 2.58    \\ 
$^{\surd}$ RXCJ\,0131.8$-$1336 & A\,209 &  01 31 53.0 &  $-$13 36 34 & 0.2060 &   6.289 
& 1.58 & 2.69    \\
$^{\surd}$ RXCJ\,0307.0$-$2840 & A\,3088 &  03 07 04.1 &  $-$28 40 14 & 0.2537 &   6.953 
& 1.69 & 2.67    \\
           RXCJ\,0437.1$+$0043 &  $-$  &  04 37 10.1 &  $+$00 43 38 & 0.2842 &   8.989 
& 2.02 & 2.79    \\
$^{\surd}$ RXCJ\,0454.1$-$1014 & A\,521 &  04 54 09.1 &  $-$10 14 19 & 0.2475 &   8.178 
& 1.89 & 2.78    \\
           RXCJ\,0510.7$-$0801 &  $-$  &  05 10 44.7 &  $-$08 01 06 & 0.2195 &   8.551 
& 1.95 & 2.86    \\
$^{\surd}$ RXCJ\,1023.8$-$2715 & A\,3444 &  10 23 50.8 &  $-$27 15 31 & 0.2542 &  13.760 
& 2.69 & 3.12    \\
$^{\surd}$ RXCJ\,1115.8$+$0129 &  $-$  &  11 15 54.0 &  $+$01 29 44 & 0.3499 &  13.579 
& 2.67 & 2.95    \\
$^{\star}$ RXCJ\,1131.9$-$1955 & A\,1300 &  11 31 56.3 &  $-$19 55 37 & 0.3075 &  13.968 
& 2.72 & 3.04    \\
           RXCJ\,1212.3$-$1816 &  $-$  &  12 12 18.9 &  $-$18 16 43 & 0.2690 &   6.197 
& 1.56 & 2.58    \\
$^{\surd}$ RXCJ\,1314.4$-$2515 &  $-$  &  13 14 28.0 &  $-$25 15 41 & 0.2439 &  10.943 
& 2.30 & 2.98    \\
$^{\surd}$ RXCJ\,1459.4$-$1811 & S\,780 &  14 59 29.3 &  $-$18 11 13 & 0.2357 &  15.531 
& 2.92 & 3.24    \\
           RXCJ\,1504.1$-$0248 &  $-$  &  15 04 07.7 &  $-$02 48 18 & 0.2153 &  28.073 
& 4.37 & 3.75    \\
$^{\surd}$ RXCJ\,1512.2$-$2254 &  $-$  &  15 12 12.6 &  $-$22 54 59 & 0.3152 &  10.186 
& 2.19 & 2.81    \\
           RXCJ\,1514.9$-$1523 &  $-$  &  15 14 58.0 &  $-$15 23 10 & 0.2226 &   7.160 
& 1.73 & 2.74    \\
$^{\star}$ RXCJ\,1615.7$-$0608 & A\,2163 &  16 15 46.9 &  $-$06 08 45 & 0.2030 &  23.170 
& 3.84 & 3.62    \\
$^{\surd}$ RXCJ\,2003.5$-$2323 &  $-$  &  20 03 30.4 &  $-$23 23 05 & 0.3171 &   9.248 
& 2.05 & 2.75    \\
           RXCJ\,2211.7$-$0350 &  $-$  &  22 11 43.4 &  $-$03 50 07 & 0.2700 &   7.418 
& 1.77 & 2.69    \\
$^{\surd}$ RXCJ\,2248.5$-$1606 & A\,2485 &  22 48 32.9 &  $-$16 06 23 & 0.2472 &   5.100 
& 1.37 & 2.50    \\ 
$^{\surd}$ RXCJ\,2308.3$-$0211 & A\,2537 &  23 08 23.2 &  $-$02 11 31 & 0.2966 &  10.174 
& 2.19 & 2.85    \\
$^{\surd}$ RXCJ\,2337.6+0016   & A\,2631 &  23 37 40.6 &  $+$00 16 36 & 0.2779 &   7.571 
& 1.79 & 2.69    \\
$^{\surd}$ RXCJ\,2341.2$-$0901 & A\,2645 &  23 41 16.8 &  $-$09 01 39 & 0.2510 &   5.789 
& 1.49 & 2.57    \\ 
$^{\surd}$ RXCJ\,2351.6$-$2605 & A\,2667 &  23 51 40.7 &  $-$26 05 01 & 0.2264 &  13.651 
& 2.68 & 3.16    \\
\noalign{\smallskip}
\hline
\end{tabular}
\end{center}
Symbols are as follows: $^{\surd}$ marks the clusters observed by us with the 
GMRT as part of our radio halo survey;  
$^{\star}$ marks the clusters with radio halo known from the literature 
(A\,2744 Govoni et al. \cite{govoni01}; A\,1300 Reid et al. \cite{reid99}; 
A\,2163 Herbig \& Birkinshaw \cite{herbig94} and Feretti et al. 
\cite{feretti01}). All the remaining clusters are part of the GMRT cluster 
Key Project (P.I. Kulkarni).
\end{table*}
%
%

In order to obtain a statistically significant sample of clusters
suitable for our aims, we based our selection on the ROSAT--ESO
Flux Limited X--ray (REFLEX) galaxy cluster catalogue 
(B{\"o}hringer et al. \cite{boeringer04}) and on the extended
ROSAT Brightest Cluster Sample (BCS)
catalogue (Ebeling et al. \cite{ebeling98} \& \cite{ebeling00}).
Here we will concentrate on the REFLEX sample, which was  
observed with the GMRT in January 2005 (present paper, see next Section), 
in October 2005 and August 2006 (Venturi et al., in preparation).
\\
From the REFLEX catalogue we selected all clusters satisfying the following 
criteria:

\begin{itemize} 
\item[1)] L$_{\rm X}$(0.1--2.4 keV) $>$ 5 $\times$ 10$^{44}$ erg s$^{-1}$;

\item[2)] 0.2 $<$ z $<$ 0.4;
 
\item[3)] $-$30$^{\circ}$ $<$ $\delta$ $<$ +2.5$^{\circ}$.
\end{itemize} 

\begin{figure}
\centering
\includegraphics[angle=0,width=8.5cm]{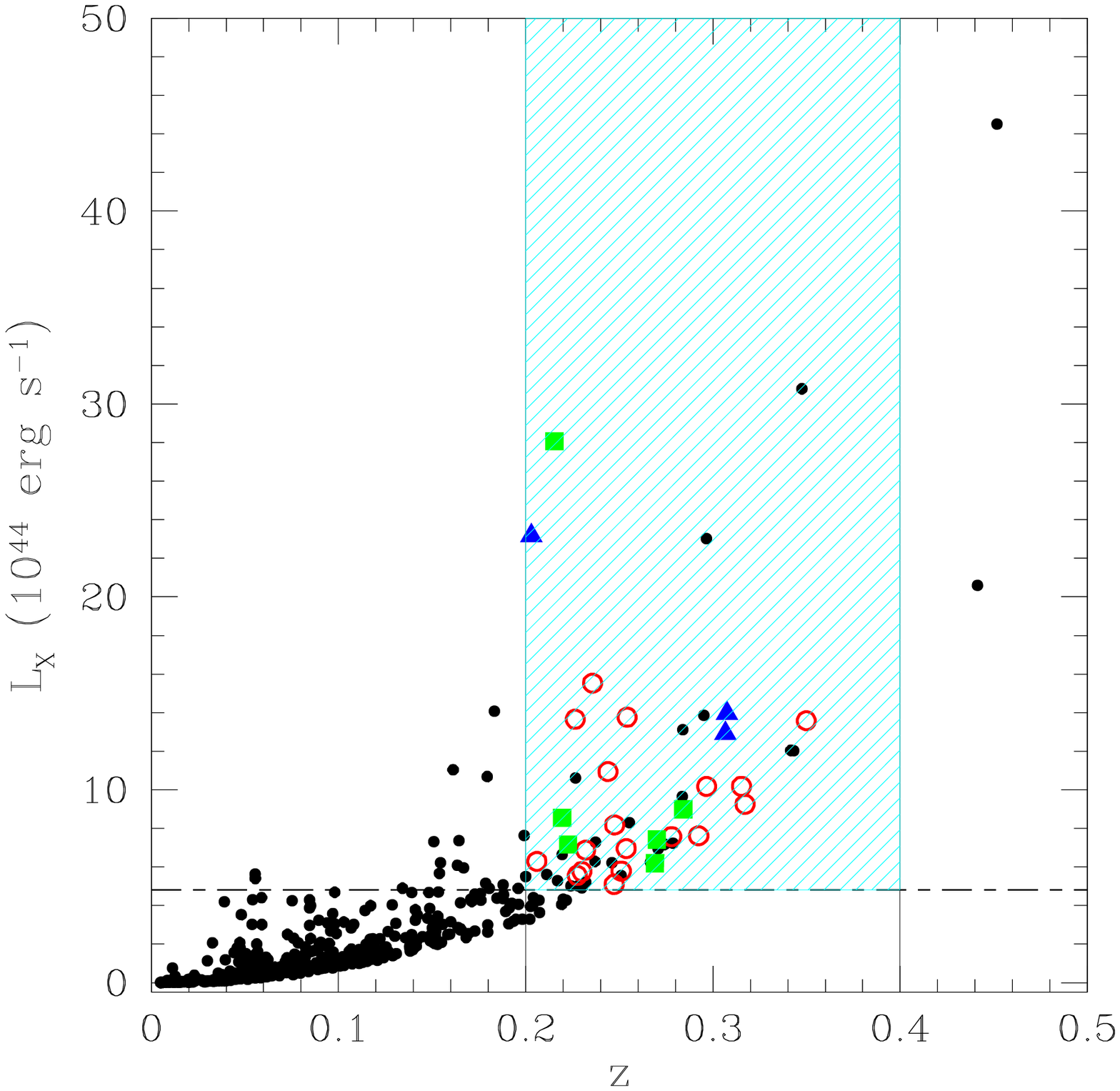}
\caption{L$_x$--z plot (0.1--2.4 KeV) for the REFLEX clusters. Open red 
circles show the clusters selected for the GMRT observations of the present 
project (marked with the symbol $\surd$ in Table 1; see Section \ref{sec:obs} 
for details); filled blue triangles indicate those clusters 
(marked with $^{\star}$
in Table 1) which are known to host a radio halo from the literature, 
i.e. A\,2744 (Govoni et al. \cite{govoni01}), A\,1300 (Reid et al. 
\cite{reid99}) and A\,2163 (Herbig \& Birkinshaw \cite{herbig94}; Feretti 
et al. \cite{feretti01}); filled green squares indicate the clusters of the 
sample belonging to the GMRT Cluster Key Project (P.I. Kulkarni). 
The light blue dashed region is the one surveyed in
our project.}
\label{fig:fzx}
\end{figure}

The lower limit of $\delta=-30^{\circ}$ was chosen 
in order to ensure a good u--v coverage with the GMRT, while the value of 
$\delta=+2.5^{\circ}$ is the REFLEX 
upper limit. 
\\
The limit in X--ray luminosity is aimed at selecting
massive clusters, which are expected to host giant radio halos
(CBS04, CB05 and references therein). It corresponds to a lower limit
in the virial mass of M$_{\rm V}~>~ 1.4\times10^{15} M_{\odot}$
if the L$_{\rm X}$ -- M$_{\rm V}$ derived in CBS06 is assumed. 
We point out that the L$_{\rm X}$ -- M$_{\rm V}$ correlation in CBS06 has 
a statistical dispersion 
of $\sim$ 30\%. This error dominates over the systematic additional
uncertainty introduced by the fact that the correlation was obtained 
using the z $<$ 0.2 cluster sample in Reiprich \& B{\"o}hringer
(\cite{reiprich02}).
\\
We obtained a total of 27 clusters. The source list is
reported in Table 1,
where we give (1) the REFLEX name, (2) alternative name from other catalogues, 
(3) and (4) J2000 coordinates, (5) redshift, (6) the X--ray luminosity in the 
0.1--2.4 keV band, (7) and (8) estimates for the virial mass M$_{\rm V}$ 
and virial radius R$_{\rm V}$ respectively (from the 
L$_{\rm X}$ -- M$_{\rm V}$ correlation derived in CBS06). 
\\
The location of the 27 clusters of the sample in the L$_{\rm X}$--z 
plane for the whole REFLEX catalogue is reported in Fig. \ref{fig:fzx}. 
%
%

\section{Radio observations}\label{sec:obs}

From the sample given in Table 1 we selected all clusters 
with no radio information available in the literature at the time
our GMRT proposal was written. We also excluded all clusters belonging
to the GMRT Cluster Key Project (P.I. Kulkarni), and remained with 
a total of 18 clusters, marked with the symbol $\surd$ in 
Table 1.
\\
From the list of marked clusters, 11 were given higher priority and
were observed with the GMRT during a 27--hour run allocated in January 2005. 
Table \ref{tab:obs} reports the following information: cluster name, 
half power beamwidth (HPWB) of the full array of the observations (arcsec), 
total time on source (minutes) and rms (1$\sigma$ in $\mu$Jy b$^{-1}$) 
in the full resolution image.
\\
Five clusters listed in Table 1 were observed with the GMRT in
a second observing run carried out in October 2005, i.e. A\,2813, A\,2485, 
A\,2895, RXCJ\,1115.8+0129 and RXCJ\,1512.2--2254; finally the two remaining 
clusters A\,2645 and A\,2667 will be observed in August 2006.
They will all be presented in a future paper (Venturi et al.
in preparation). 
\\
%
\begin{table*} 
\caption[]{GMRT observations.}
\begin{center}
\begin{tabular}{lccc}
\hline\noalign{\smallskip}
Cluster  & Beam, PA  & Obs. time    &   rms      \\ 
          &  (full array) $^{\prime \prime} \times^{\prime \prime}$, $^{\circ}$&
min &  $\mu$Jy b$^{-1}$\\
\noalign{\smallskip}
\hline\noalign{\smallskip}
A\,2697             &  8.5$\times$5.0, --83 &  90  &  80   \\
A\,141              &  7.7$\times$7.4, 75   & 150  &  100  \\ 
A\,209              &  8.0$\times$5.0, 64   & 240  &  60   \\
A\,3088             &  8.0$\times$7.0, 40   & 190  &  65   \\ 
A\,521              &  8.6$\times$4.0, 57   & 210  &  35   \\
A\,3444             &  7.6$\times$4.9, 19   & 120  &  67   \\
RXCJ\,1314.4--2515  &  8.0$\times$5.0  15   & 150  &  65   \\
S\,780              &  7.5$\times$5.0  25   &  80  &  70   \\
RXCJ\,2003.5--2323  &  6.9$\times$5.0, --3  & 240  &  40   \\
A\,2537             &  10.3$\times$6.0  67  & 150  &  60   \\
A\,2631             &  9.2$\times$6.3 --77  & 240  &  50   \\
\hline{\smallskip}
\end{tabular}
\end{center}
\label{tab:obs}
\end{table*}

The observations were carried out at 610 MHz, using simultaneously
two 16 MHz bands (upper side band, USB, and lower side band, LSB), 
for a total of 32 MHz. Left and
right polarization were recorded for each band.
The observations
were carried out in spectral line mode, with 128 channels each
band, and a spectral resolution of 125 kHz/channel.
The data reduction and analysis were carried out with the NRAO
Astronomical Image Processing System (AIPS) package.
In order to reduce the size of the dataset, after bandpass calibration
the central 94 channels were averaged to 6 channels of $\sim$ 2 MHz each.
For each source the USB and LSB datasets, as well as the datasets
taken in different days, were calibrated and reduced separately,
then the final images from each individual dataset were combined in the 
image plane to obtain the final image. Wide--field imaging  
was adopted in each step of the data reduction.
\\
For each cluster we produced images over a wide range of resolutions,
in order to fully exploit the information GMRT can provide.
We point out that the nominal largest detectable structure provided
by the GMRT at 610 MHz is 17$^{\prime}$. This value ensures the possible 
detection of the extended radio sources we are searching for, 
since  the angular scale covered by a 1 Mpc--size structure is 
$\sim~5^{\prime}$ at z=0.2 and $\sim~3^{\prime}$ at z=0.4.
\\
The sensitivity of our
observations (1$\sigma$ level) is in the range 35 -- 100 $\mu$Jy for
the full resolution images (see Table \ref{tab:obs}), which were
obtained by means of uniform weighting.
The spread in the noise level depends most critically 
on the total time on source, on the total bandwidth available (in few 
cases only one portion of the band provided useful data, 
see individual clusters in \textsection \ref{sec:results}), 
and on the presence of strong sources in  the imaged field.
Slightly lower values for the noise level are obtained for the low resolution
images (see Section \ref{sec:results} and figure captions), which were made
using natural weighting. 
\\
The average residual amplitude errors in our data are of the order of 
\ltsim$~$5\%.
\\

\section{Results}\label{sec:results}


\medskip\noindent
\begin{table*}[t] 
\caption[]{Parameters of the extended cluster radio sources.}
\begin{center}
\begin{tabular}{lcrclrl}
\hline\noalign{\smallskip}
Cluster & Source Type & S$_{\rm 610~MHz}$ & logP$_{\rm 610~MHz}$ &  LAS & LLS & L$_1$/L$_2$    \\ 
        &             & mJy               &  W Hz$^{-1}$         & arcmin & kpc &     \\
\noalign{\smallskip}
\hline\noalign{\smallskip}
A\,209           & Giant Halo &   24.0 $\pm$ 3.6  &  24.46  & $\sim$ 4 & $\sim$  810 & $\sim$ 2   \\
A\,521           & Relic      &   41.9 $\pm$ 2.1  &  24.91  & $\sim$ 4 & $\sim$  930 & $\sim$ 4.5 \\ 
RXCJ\,1314.4--2515  & Western Relic &   64.8 $\pm$ 3.2 &  25.03 & $\sim$ 4   & $\sim$  910 & $\sim$ 3 \\
                  & Eastern Relic &   28.0 $\pm$ 1.4 &  24.67 & $\sim$ 4   & $\sim$  910 & $\sim$ 4.3 \\
                  & Halo          &   10.3 $\pm$ 0.3  &  24.22  & $\sim$ 2 & $\sim$  460 & $\sim$ 1.5 \\
RXCJ\,2003.5--2323  & Giant Halo  &   96.9 $\pm$ 5.0  &  25.49  & $\sim$ 5   & $\sim$ 1400 & $\sim$ 1.3 \\
A\,3444   & Central Galaxy        &   16.5 $\pm$ 0.8  &  24.51  & $\sim$0.7  & $\sim$  165 & $\sim$ 1.4  \\
          & surrounding Halo      &   10.0 $\pm$ 0.8 &   24.29  & $\sim$1.5  & $\sim$  350 & $\sim$ 1.4  \\
\hline{\smallskip}
\end{tabular}
\end{center}
\label{tab:param}
\end{table*}
\\
Cluster scale radio emission either in the form of radio halo or relic
was detected in 4 clusters of the sample (\textsection \ref{sec:halos}); 
in one cluster extended emission was found around the dominant 
galaxy (\textsection \ref{sec:minihalo});
for the remaining six clusters no hint of extended emission is present at
the sensitivity level of the observations (\textsection \ref{sec:noext}). 
Details on each cluster are given in this Section. In Appendix A we report the
610 MHz radio contours within the virial radius for all the observed clusters.
All the images were convolved with a HPWB of 15.0$^{\prime \prime} \times
12.0^{\prime \prime}$, except  for A\.209, RXCJ\,1314.4--2515 and
RXCJ\,2003.5--2323 where a different resolution was chosen in 
order to complement the information provided in Figs. 2, 3, 4, 5 and 6.
Table \ref{tab:param}  reports the 
observational information for the detected cluster--scale radio 
sources. The last column in the table, L$_1$/L$_2$, provides the ratio 
between the major (LAS) and minor axis of the extended emission.
The linear size and flux densities were derived from the
3$\sigma$ contour level.

%
%
\begin{figure*}
\hspace{0.5truecm}\includegraphics[angle=0,width=7.6cm, height=6.5cm]{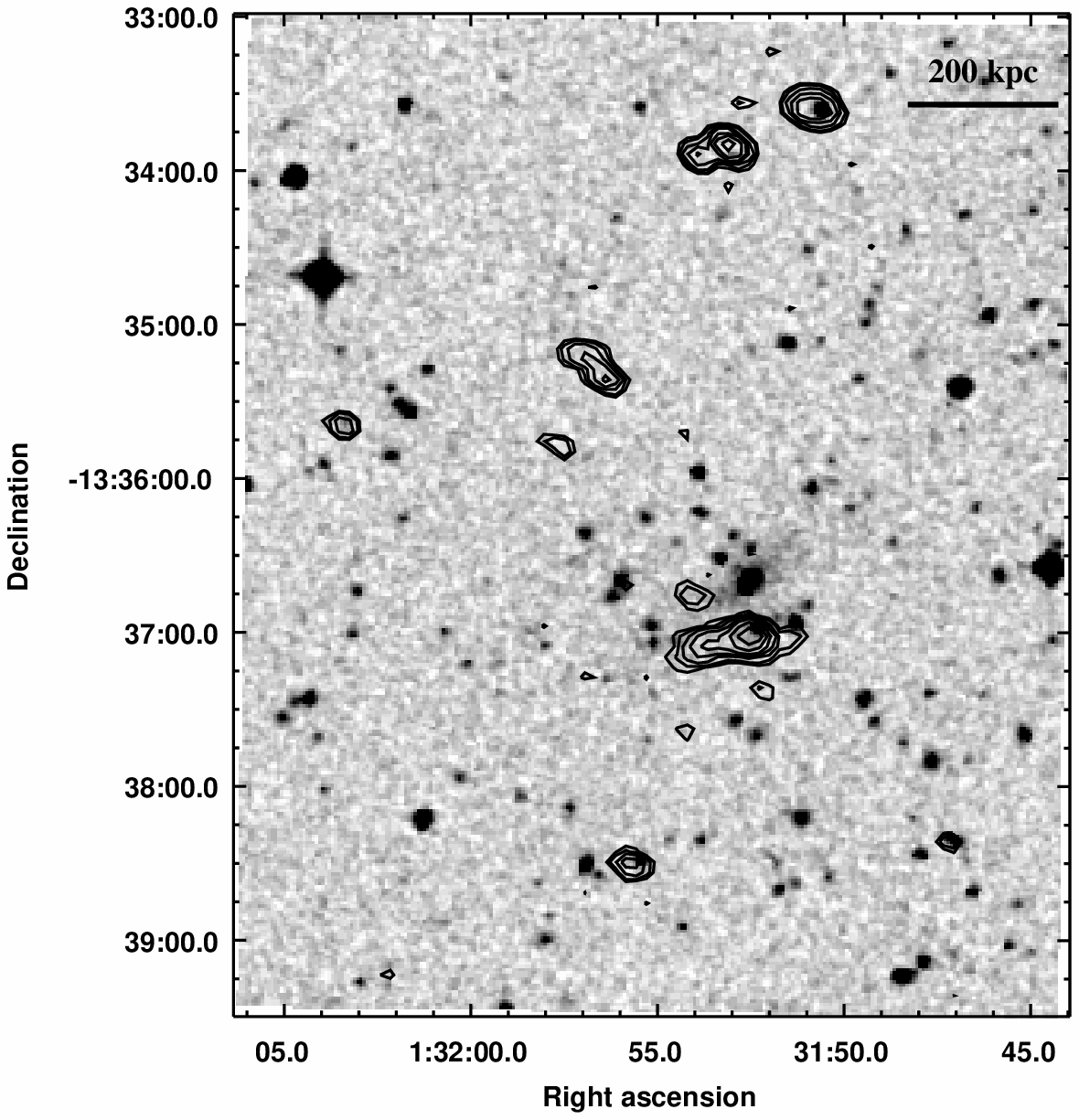}
\hspace{1.5truecm}\includegraphics[angle=0,width=7.2cm, height=6.5cm]{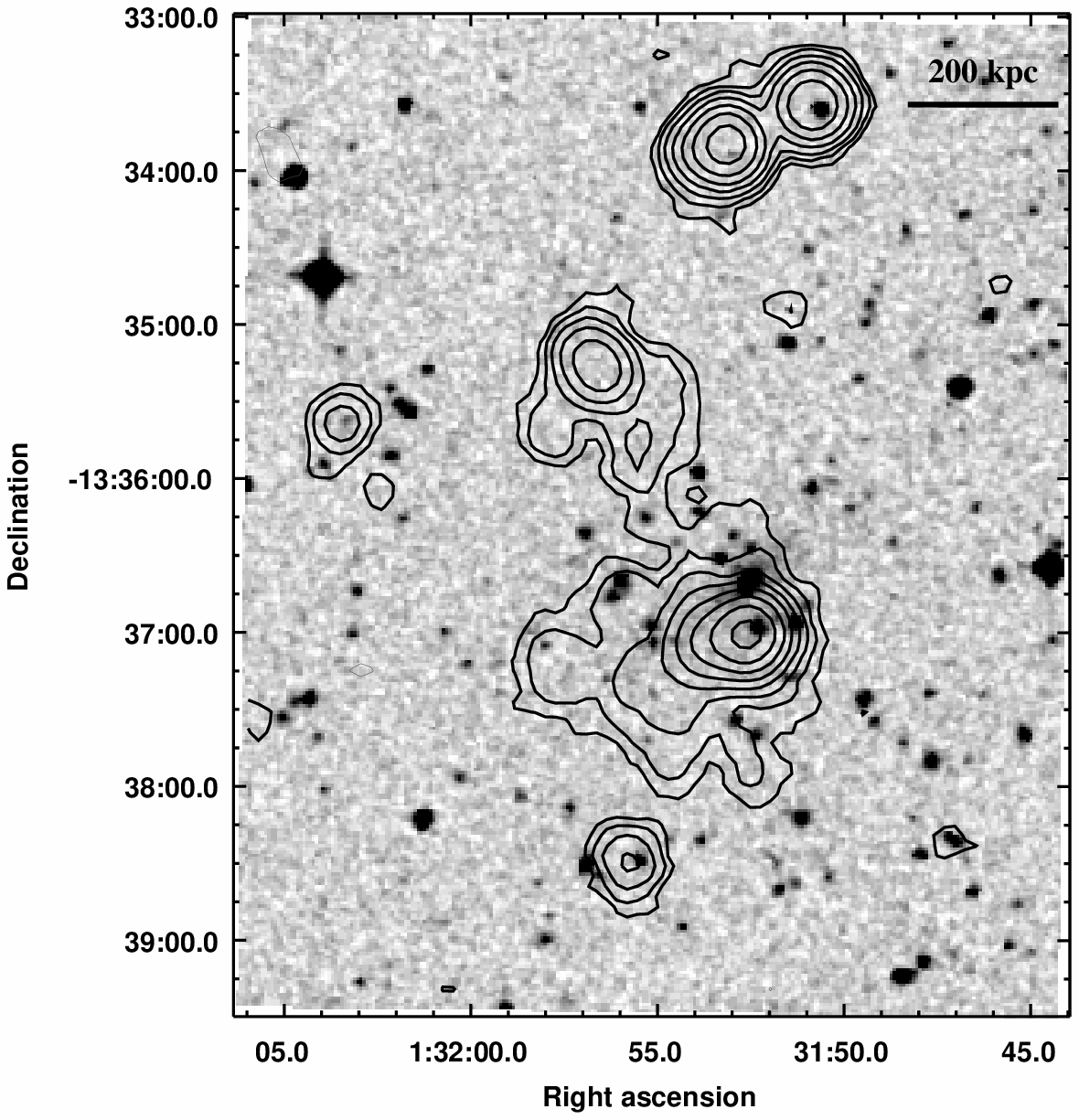}
\caption{Left -- GMRT 610 MHz radio contours for the A\,209 cluster
superposed on the POSS--2 optical plate.
The 1$\sigma$ level in the image is 60 $\mu$Jy b$^{-1}$. Contours are 
0.3$\times(\pm$ 1,2,4,8,16...) mJy b$^{-1}$. The 
HPWB is $8.0^{\prime\prime} \times 5.0^{\prime\prime}$, p.a.
$64^{\circ}$. Right -- Natural weighted image of the same sky region at 
the resolution of $18.0^{\prime\prime} \times 17.0^{\prime\prime}$, p.a.
$0^{\circ}$. The rms (1$\sigma$) in the image is 60 $\mu$Jy b$^{-1}$,
contours are 0.18$\times(\pm$ 1,2,4,8,16...) mJy b$^{-1}$.}
\label{fig:a209_opt}
\end{figure*}

\subsection{Clusters with halos, giant halos and relics}\label{sec:halos}

\subsubsection{Abell 209}

\begin{figure}
\hspace{0.8truecm}\includegraphics[angle=0,width=7.2cm, height=6.7cm]{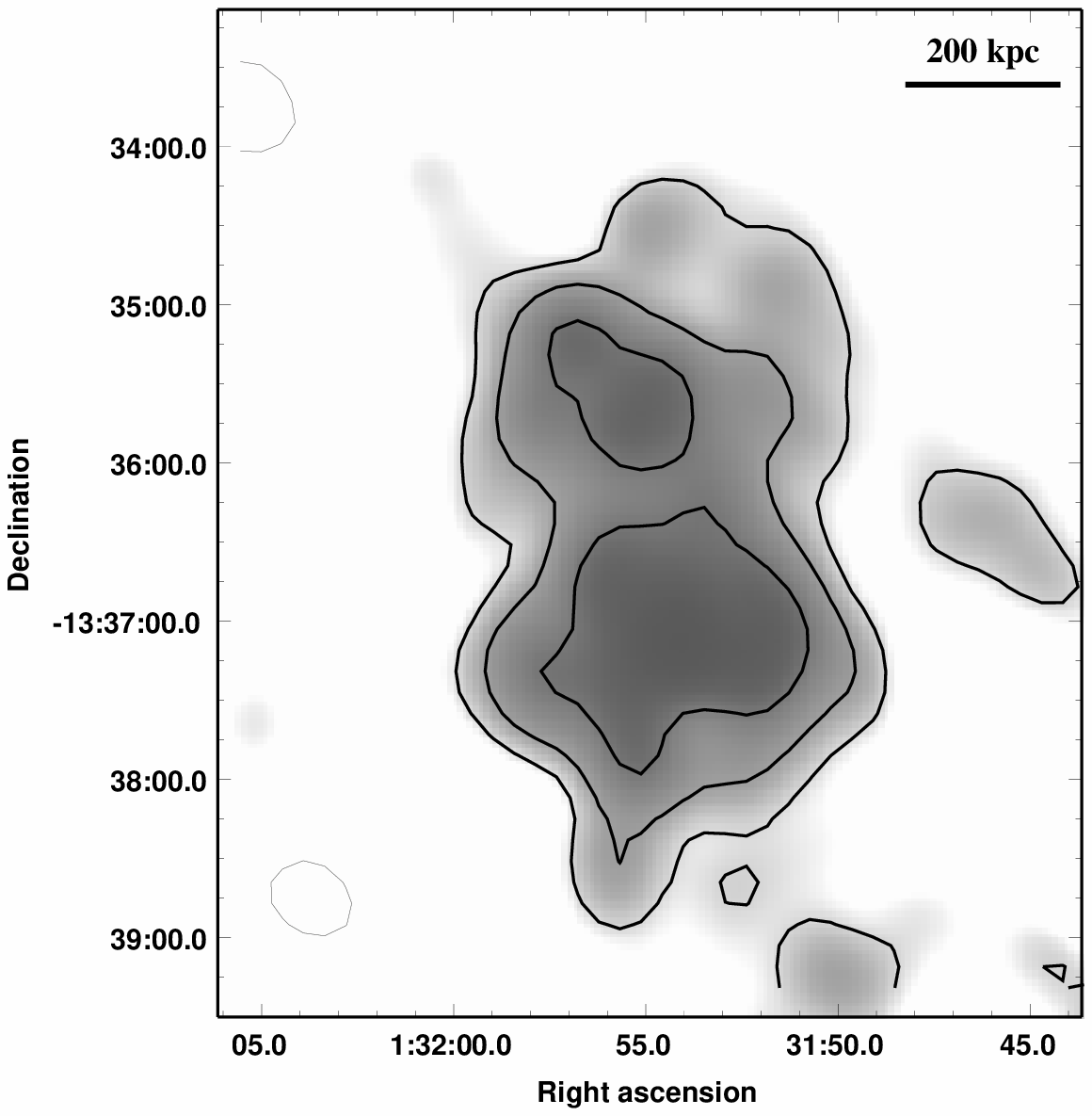}
\caption{Radio contours over grey scale of the A\,209 cluster 
radio halo after subtraction of the individual radio sources (see left panel
of Fig. \ref{fig:a209_opt} and Sect. 4.1.1 in the text). The resolution of this 
image is $32.0^{\prime\prime} \times 30.0^{\prime\prime}$,
p.a. $30^{\circ}$. The rms (1$\sigma$) in the image is 0.15 mJy b$^{-1}$,
contours are 0.35$\times(\pm$ 1,2,4,8,16...) mJy b$^{-1}$.} 
\label{fig:a209_halo}
\end{figure}

Abell 209 (RXCJ\,0131.8--1336) is a richness R=3 cluster at 
z=0.2060 (1$^{\prime\prime}$=3.377 kpc). A high 
X--ray temperature is reported in the literature.
Rizza et al. (\cite{rizza98}) estimated a mean gas 
temperature of kT$\sim$10 keV from the ROSAT 
X--ray luminosity; this  
value was confirmed by Mercurio et al. (\cite{mercurio04a}) 
from the analysis of {\it Chandra} archive data. 
\\
The cluster has been extensively studied at optical (Mercurio et al.
\cite{mercurio03}, Mercurio et al. \cite{mercurio04a} and \cite{mercurio04b},
Haines et al.  \cite{haines04}), and X--ray wavelenghts (ROSAT--HRI, 
Rizza et al. \cite{rizza98}, Mercurio et al. \cite{mercurio04a}). 
These studies show that A\,209 is far from a relaxed dynamical stage, and 
it is undergoing a strong dynamical evolution. 
In particular the X--ray and the optical data suggest that A\,209 is 
experiencing a merging event between two or more components.
\\
Mercurio et al. (\cite{mercurio03}) provided an estimate of the virial
mass of the cluster, $M_{\rm V} = 2.25^{+0.63}_{-0.65}\times 10^{15}M_{\odot}$,
consistent with our estimate given in Table 1 if we account for the
uncertainty of our value (see Sect. \ref{sec:sample}).
\\
The cluster merging scenario is confirmed by the weak lensing analysis 
carried out by Dahle et al. (\cite{dahle02}), who found two significant peaks
in the mass distribution of the cluster: the largest one is close to the central 
cD galaxy, and the secondary mass peak is located at about 5 arcmin 
north of the cluster centre and associated to a peak in the optical galaxy
distribution.
\\
\\
610 MHz contours of the A\,209 emission within the virial radius  
are given in Fig. \ref{fig:a209_lr}, while  Fig. \ref{fig:a209_opt} 
shows the central part of the field at two different resolutions 
superposed on the POSS--2 image. Inspection of Fig. \ref{fig:a209_lr}
and of the right panel in  Fig. \ref{fig:a209_opt} 
coupled with flux density measurements, suggests the presence
of extended emission around the individual central cluster 
radio galaxies. 
\\
In order to highlight such emission we subtracted all the individual 
sources visible in the full resolution image from the u--v data, and 
convolved the residuals with a 
HPBW with size $32.0^{\prime\prime} \times 30.0^{\prime\prime}$. 
The image is reported in Fig. \ref{fig:a209_halo}.
%
%
The adopted procedure indeed confirms the existence of cluster 
scale extended emission. The possible presence of a radio halo in 
A\,209 was suggested by Giovannini et al. (\cite{giovannini99}) 
from inspection of the NRAO VLA Sky Survey (NVSS), and confirmed
in Giovannini et al. (\cite{gg06}) on the basis of 1.4 GHz VLA observations.
Our GMRT image in Fig. \ref{fig:a209_halo}  
is in partial agreement with the size and morphology of the VLA 1.4 GHz 
image shown by those authors. The largest angular size (LAS) 
is $\sim 4^{\prime}$, i.e. $\sim$ 810 kpc, therefore
we classify the source as a {\it giant} radio halo. Its total 
flux density, measured after subtraction of the individual radio sources 
(see left panel of Fig. \ref{fig:a209_opt}) is 
S$_{\rm 610~MHz} = 24.0 \pm 3.6$ mJy, which implies a total radio power of 
logP$_{\rm 610~MHz}$ (W/Hz)= 24.46.
The difficulty in subtracting the extended individual sources (in particular 
the head--tail radio galaxy located just South of the cluster centre) reflects 
both in the large error associated with the flux density measurement, 
and in the unusual brightness distribution of the radio halo, characterised 
by two peaks of emission.
\\
Further observations are already in progress with the GMRT, in
order to better image and study this source.

\subsubsection{Abell 521}

A detailed study of A\,521 (RXCJ\,0454.1--1014, z=0.2475, 
1$^{\prime\prime}$=3.875 kpc) has already been published by Giacintucci et
al. (\cite{giacintucci06}). This merging cluster hosts a 
radio relic located at the border of the X--ray emission. 
We discussed the origin of this source in the light of current scenarios 
for the formation of radio relics, i.e. 
acceleration of electrons from the thermal pool or compression
of fossil radio plasma, both through merger shock waves. 
We refer to that paper for the images and radio information 
and will include A\,521 in the discussion in Section \ref{sec:discussion}.
All values and observational parameters reported in 
Table \ref{tab:obs} and \ref{tab:param} are taken from Giacintucci
et al. (\cite{giacintucci06}).

\subsubsection{RXCJ\,1314.4$-$2515}

%
%
\begin{figure*}
\centering
\hspace{-0.5truecm}\includegraphics[angle=0,width=7.4cm, height=6.4cm]{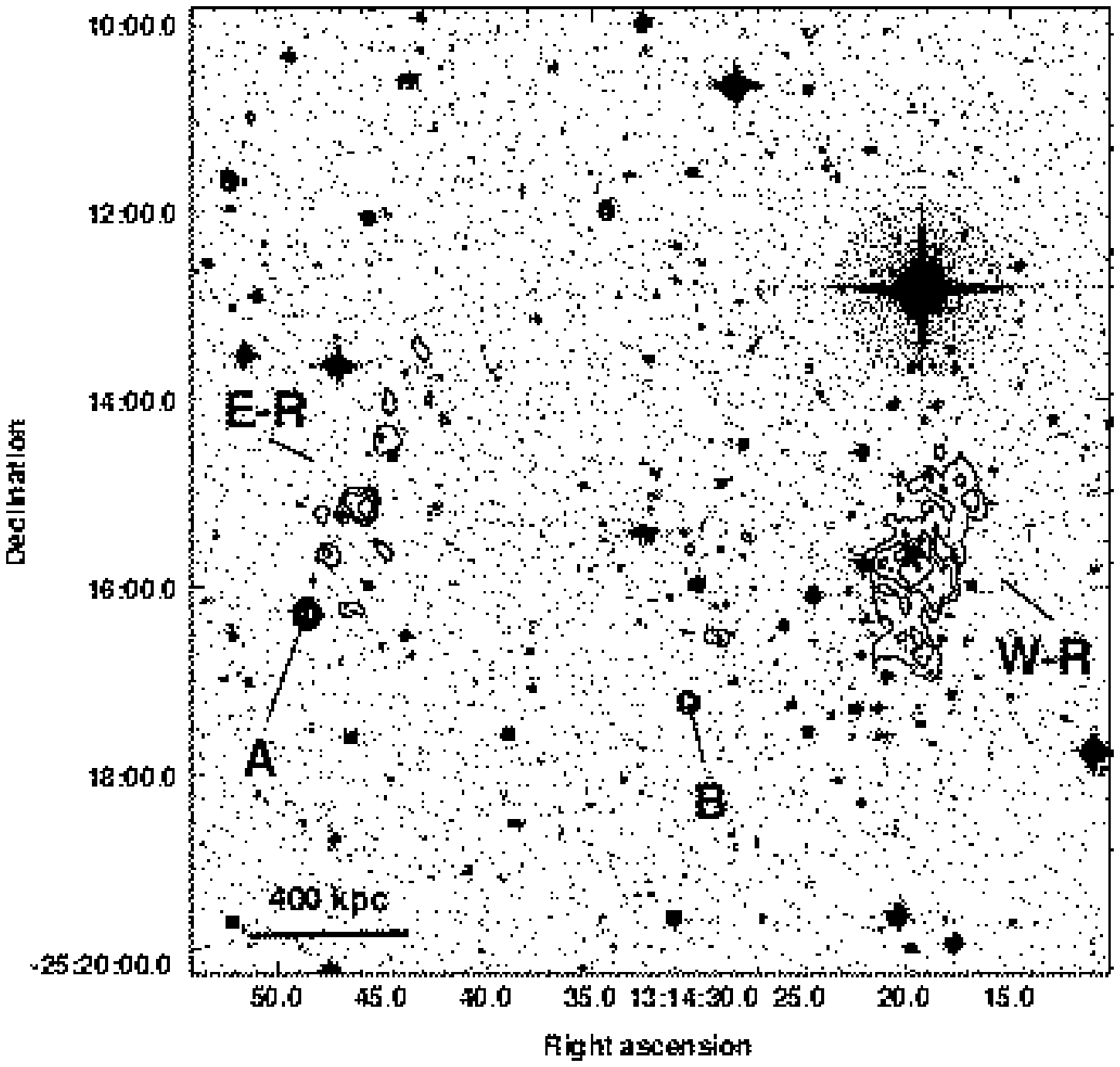}
\hspace{1truecm}\includegraphics[angle=0,width=8cm, height=6.5cm]{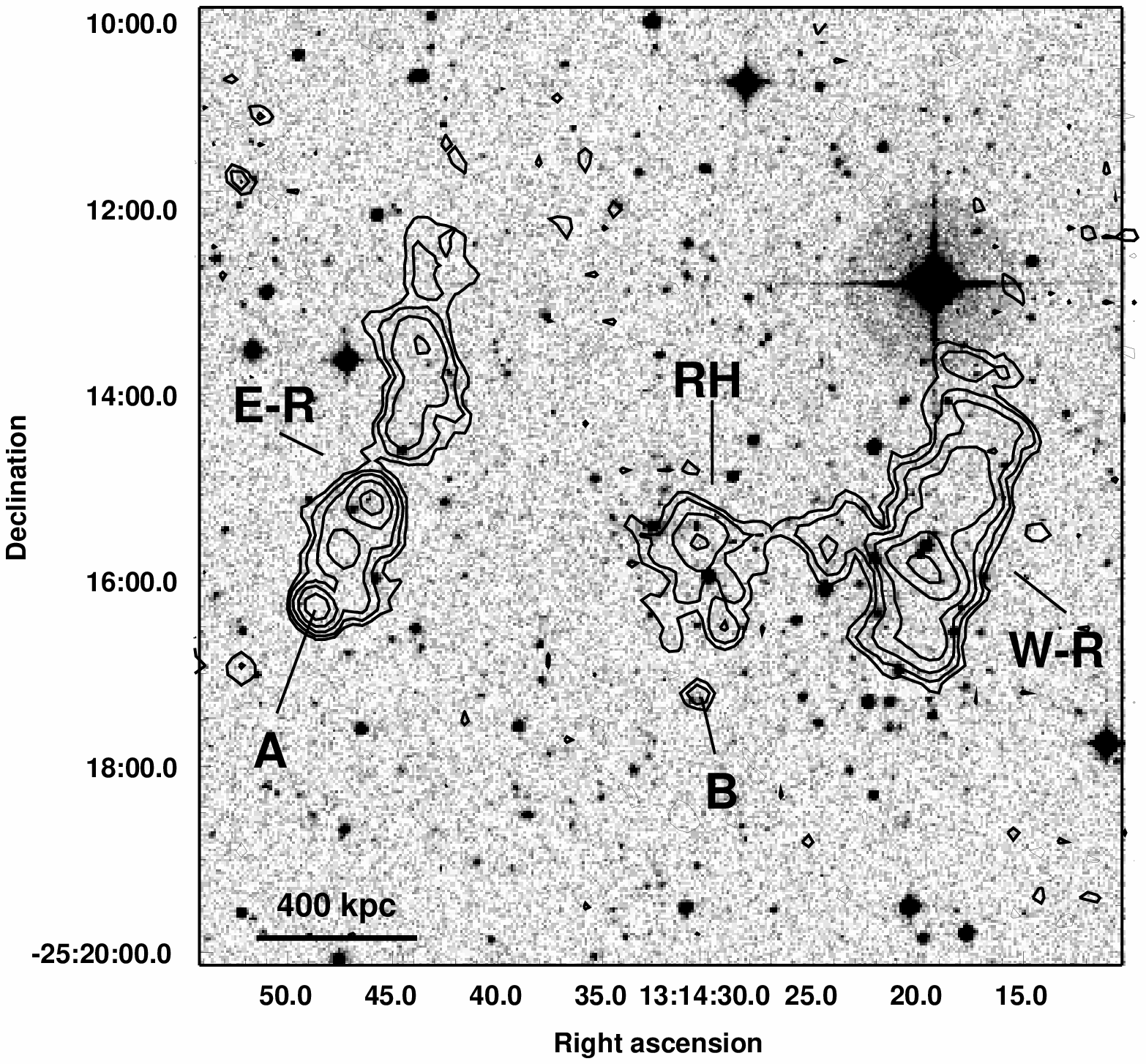}
\caption{Left --
GMRT 610 MHz radio contours for the cluster RXCJ\,1314.4--2515
superposed on the POSS--2 optical plate.
The 1$\sigma$ level in the image is 60 $\mu$Jy b$^{-1}$.
Contours are 0.18$\times(\pm$1,2,4,8,16...)  mJy b$^{-1}$. The 
HPWB is $8.0^{\prime\prime} \times 5.0^{\prime\prime}$, p.a.
$15^{\circ}$. The western and the eastern relics are labelled 
as E--R and W--R respectively, and the individual point sources 
in the relics/halo region are indicated as A and B.
Right -- GMRT 610 MHz radio contours for the cluster RXCJ\,1314.4--2515
superposed on the POSS--2 optical plate.
The 1$\sigma$ level in the image is 60 $\mu$Jy b$^{-1}$.  
Contours are 0.2$\times(\pm$1,2,4,8,16...) mJy b$^{-1}$. The 
HPWB is $20.0^{\prime\prime} \times 15.0^{\prime\prime}$, p.a.
$39^{\circ}$ The western and the eastern relics are labelled 
as E--R and W--R respectively, RH indicates the radio halo, and
the individual point sources in the relics/halo region are 
indicated as A and B.}
\label{fig:rxcj1314_rel2}
\end{figure*}

Evidence of a disturbed dynamical status for the cluster 
RXCJ\,1314.4$-$2515 (z=0.2439, 1$^{\prime\prime}$=3.806 kpc) 
is reported in the literature. The redshift distribution of the 
cluster galaxies clearly shows a bimodal structure, 
with two peaks separated in velocity space by $\sim$ 1700 km s$^{-1}$ 
(Valtchanov et al. \cite{valtchanov02}). The X--ray morphology of the 
cluster is also bimodal, and it is elongated along the 
E--W direction, the western peak being the brightest 
(Valtchanov et al. \cite{valtchanov02}). 
\\
This cluster was observed with the VLA at 1.4 GHz by Feretti 
et al. (\cite{feretti05}), who revealed the presence of a radio halo at the
cluster centre and two peripheral sources, which they classified as relics.
\\
Fig. \ref{fig:rxcj1314_lr} reports the contour image of our 610 MHz 
observations at the resolution of $15.0^{\prime\prime} 
\times 13.0^{\prime\prime}$ within the virial radius. The central part 
of the cluster is given in the left and right panel of 
Fig. \ref{fig:rxcj1314_rel2}, both superposed on the POSS--2 plate.
The left panel shows the full resolution image, 
while in the right panel lower resolution contours are displayed.
Fig. \ref{fig:rxcj1314_halo_asca} shows the same region 
overlaid on the X--ray ASCA image.
Our images confirm that RXCJ\,1314.4--2515 has a complex radio morphology,
with the presence of three different regions of extended emission on the cluster
scale.
%
%
\begin{figure}\label{fig:rxcj1314_halo_asca}
\centering
\hspace{-0.5truecm}\includegraphics[angle=0,width=8.0cm]{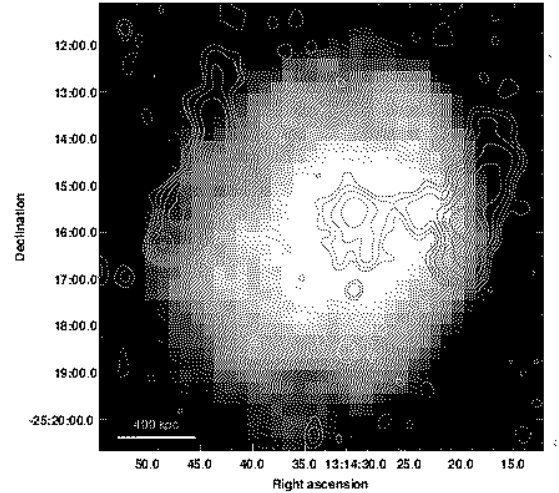}
\caption{GMRT 610 MHz radio contours for the cluster 
RXCJ\,1314.4--2515 superposed on the X--ray archive ASCA image (colour).
The 1$\sigma$ level in the image is 60 $\mu$Jy b$^{-1}$. 
Contours are 0.18$\times(\pm$1,2,4,8,16...) mJy b$^{-1}$. The 
HPWB is $25.0^{\prime\prime} \times 22.0^{\prime\prime}$, p.a.
$15^{\circ}$.} 
\label{fig:rxcj1314_halo_asca}
\end{figure}
\\
Two parallel features are easily visible in Figs. \ref{fig:rxcj1314_lr},
\ref{fig:rxcj1314_rel2} and \ref{fig:rxcj1314_halo_asca}. 
They are separated by $\sim 6^{\prime}$ and extend in the  SE--NW direction 
for approximately 4$^{\prime}$ (i.e. $\sim$ 910 kpc at the cluster redshift).
The remarkable superposition of the low resolution radio image with the  
ASCA image in Fig. \ref{fig:rxcj1314_halo_asca} clearly 
shows that these two sources are located at the border of the detected  
X--ray emission.
\\
The overall morphology of these two features, coupled with their 
location with respect to the intracluster gas, suggest that they
are radio relics, as also discussed in Feretti et al. (\cite{feretti05}), 
who ruled out any association with individual galaxies.
In the following we will 
refer to the eastern and the western relics as E--R and 
W--R respectively, as also labelled in Fig. 
\ref{fig:rxcj1314_rel2}.
The morphology and flux density ratio of the two relics are 
consistent with the 1.4 GHz data in Feretti et al. (\cite{feretti05}).
Their flux densities at 610 MHz are S$_{\rm 610~MHz} = 64.8 \pm 3.2$  mJy and 
S$_{\rm 610 MHz} = 28.0 \pm 1.4$ mJy for W--R and E--R respectively. 
The value given for E--R does not include the southernmost pointlike source 
A (Fig. \ref{fig:rxcj1314_rel2}). 
In order to derive the total spectral index of W--R and E--R between 
1.4 GHz and 610 MHz we included also the contribution of source A to the 
flux density of
E--R, for a consistent comparison with Feretti et al. (\cite{feretti05}), and
obtained 32.8 mJy. Our flux density measurements lead to the same 
value for the spectral index in both features. In particular, 
$\alpha_{\rm 610~MHz}^{\rm 1.4~GHz}$(W--R) = 1.40 $\pm$ 0.09 and
$\alpha_{\rm 610~MHz}^{\rm 1.4~GHz}$(E--R) = 1.41 $\pm$ 0.09.
\\
\\
Figs. \ref{fig:rxcj1314_rel2} (right panel)
and \ref{fig:rxcj1314_halo_asca} 
show that extended emission is present in the region between 
W--R and E--R, consistent with the 1.4 GHz VLA images in Feretti 
et al. (\cite{feretti05}), who classified
this feature as a radio halo. This source is referred to as RH in the right
panel of Fig. \ref{fig:rxcj1314_rel2}. It is spatially coincident with the 
bulk of the optical galaxies (see Valtchanov et al. \cite{valtchanov02}) and 
its largest angular size is $\sim~2^{\prime}$, corresponding to 460 kpc, 
i.e. it is not a giant radio halo.
The radio halo seems to blend with the emission of the western relic 
W--R, however it is difficult to say whether this is a true feature, since
projection effects are likely to play a role. Given the different 
polarisation properties of radio halos and relics, polarisation information
would be necessary to investigate this issue.
We measured a flux density of S$_{\rm 610~MHz} = 10.3 \pm 0.3$ mJy for the
radio halo. No spectral index estimate between 610 MHz and 1.4 GHz can be 
derived, due to the lack of a flux density value
at 1.4 GHz (Feretti et al. \cite{feretti05}). 
\\
\\

\subsubsection{RXCJ\,2003.5--2323}\label{sec:rxcj2003}

%
%
\begin{figure*}
\centering
\hspace{-1truecm}\includegraphics[angle=0,width=8.1cm, height=6.6cm]{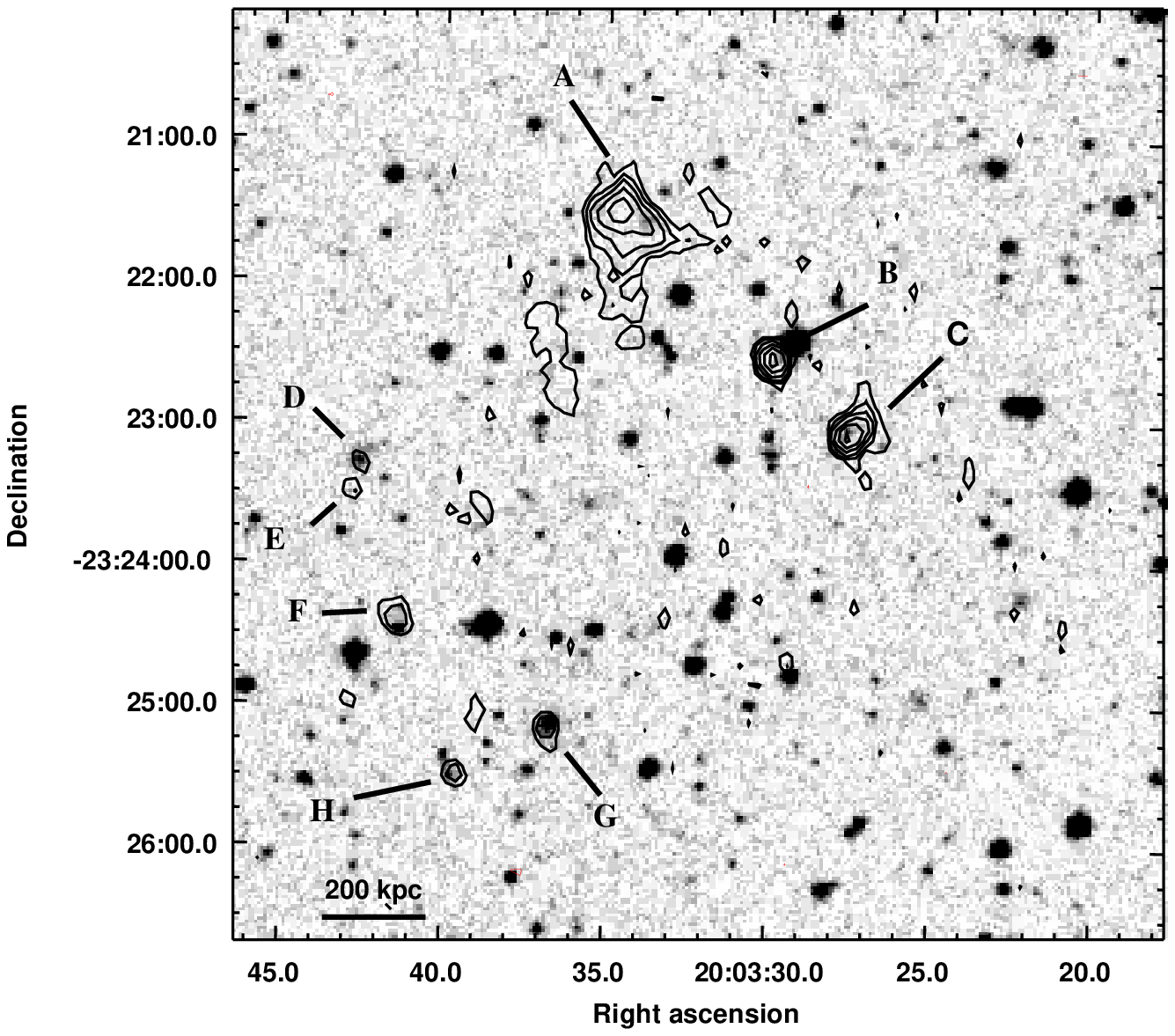}
\hspace{1truecm}\includegraphics[angle=0,width=8cm, height=6.7cm]{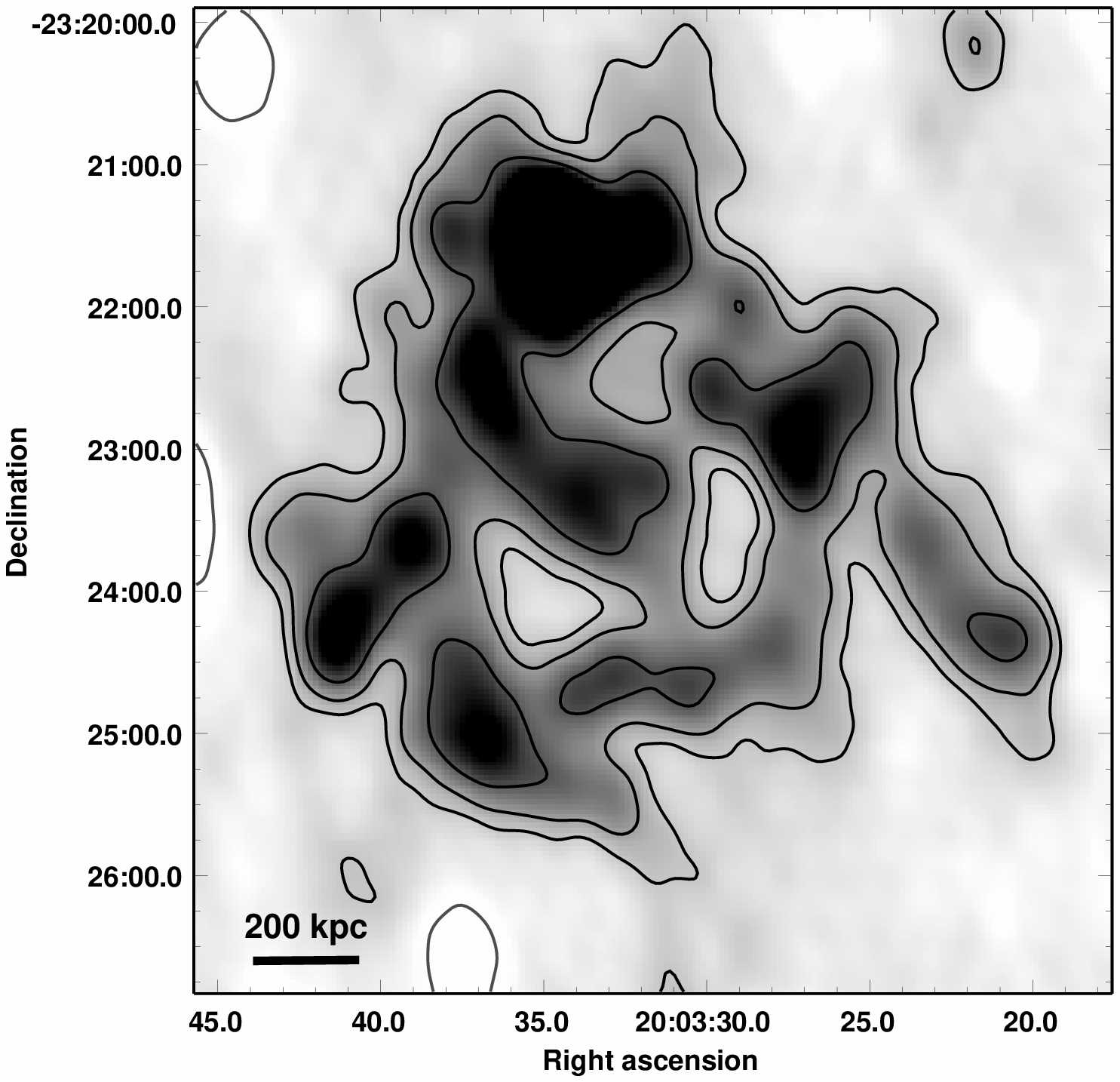}
\caption{Left -- Full resolution GMRT 610 MHz contours of the
central region of RXCJ\,2003.5--2323, superposed to the POSS--2 
optical image. The resolution of the radio image is 
$6.9^{\prime\prime} \times 5.0^{\prime\prime}$, p.a. $-0.3^{\circ}$, 
the 1$\sigma$ level is 40 $\mu$Jy b$^{-1}$. Contours are 
0.12$\times(\pm$1,2,4,8...) mJy b$^{-1}$. 
Individual sources are labelled from A to H.
Right -- GMRT 610 MHz gray scale and radio contours of the 
giant radio halo in RXCJ\,2003.5--2323 after subtraction of the 
individual sources (from B to H in the left panel). The 
HPWB is $32.0^{\prime\prime} \times 23.0^{\prime\prime}$, p.a.
$15^{\circ}$. Contours are 0.3$\times(\pm$1,2,4,8...) mJy b$^{-1}$. 
The 1$\sigma$ level in the image is 100 $\mu$Jy b$^{-1}$.}
\label{fig:rxcj2003_halo}
\end{figure*}
%
%
%

RXCJ2003.5--2323 is the most distant cluster in our 
sample, with z=0.3171 (1$^{\prime\prime}$=4.626 kpc).
Little information is available in the literature. 
The ROSAT All Sky Survey (RASS) image shows that the X--ray emission
is elongated along the NW--SE direction, which
might suggest a disturbed dynamical status for RXCJ\,2003.5--2323.
\\
Our GMRT 610 MHz observations show that it is the most striking
cluster among those observed thus far.
It hosts a {\it giant} radio halo, one of the largest known up to date.
Its largest angular size is $\sim 5^{\prime}$, corresponding to $\sim$ 
1.4 Mpc.
Hints of the presence of this very extended radio halo were clear
already from inspection of the NRAO VLA Sky Survey (NVSS).
\\
The cluster radio emission within the cluster virial radius  
is given in  Fig. \ref{fig:rxcj2003}. The central part of the cluster is 
shown in Fig. \ref{fig:rxcj2003_halo}. 
The left panel shows a full resolution image 
superposed to the POSS--2 optical image, to highlight the individual sources 
(labelled from A to H). The sources with a clear optical counterpart (B to H) were 
subtracted from the u--v data when producing the image shown in the right panel
of Fig. \ref{fig:rxcj2003_halo}, which we convolved with a larger beam 
in order to highlight the low surface brightness emission. We did not subtract
A, since no optical counterpart is visible on the POSS--2, therefore we
consider this feature as a peak in the radio halo emission.
One of the most striking features of this giant radio halo 
is its complex morphology: 
clumps and filaments are visible on angular scales of the
order of $\sim 1^{\prime}$ (clumps) and $\sim 2-3^{\prime}$ (filaments),
as clear from Figures \ref{fig:rxcj2003} and 
\ref{fig:rxcj2003_halo} (right panel).
Unfortunately no deep X--ray images are available for this cluster, 
therefore a combined radio and X--ray analysis cannot be carried out.
The only information we can derive from the RASS image of the cluster is 
that the whole radio emission from the halo is embedded within the X--ray 
emission, as shown in Fig. \ref{fig:rxcj2003_rass}.
\\
The total flux density of the radio halo (after subtraction of the point 
sources) is S$_{\rm 610~MHz} = 96.9 \pm 5.0$ mJy, corresponding to 
logP$_{\rm 610~MHz}$ (W/Hz) = 25.49.
%
%

\begin{figure}\label{fig:rxcj2003_rass}
\centering
\hspace{0.5truecm}\includegraphics[angle=0,width=6.5cm]{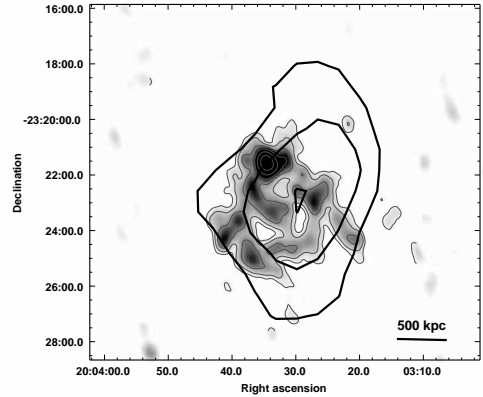}
\caption{ROSAT All Sky Survey contours (black) of RXCJ\,2003.5--2323 overlaid 
on the 610 MHz gray scale and contours (gray) of the radio halo. 
The X--ray contours levels are logarithmically spaced by a factor of 
$\sqrt 2$. The radio image is the same as right 
panel of Fig. \ref{fig:rxcj2003_halo}. }
\label{fig:rxcj2003_rass}
\end{figure}
%

\subsection{Candidate extended emission in Abell 3444}\label{sec:minihalo}

\begin{figure*}
\centering
\hspace{-0.5truecm}\includegraphics[angle=0,width=9.8cm,height=7.5cm]{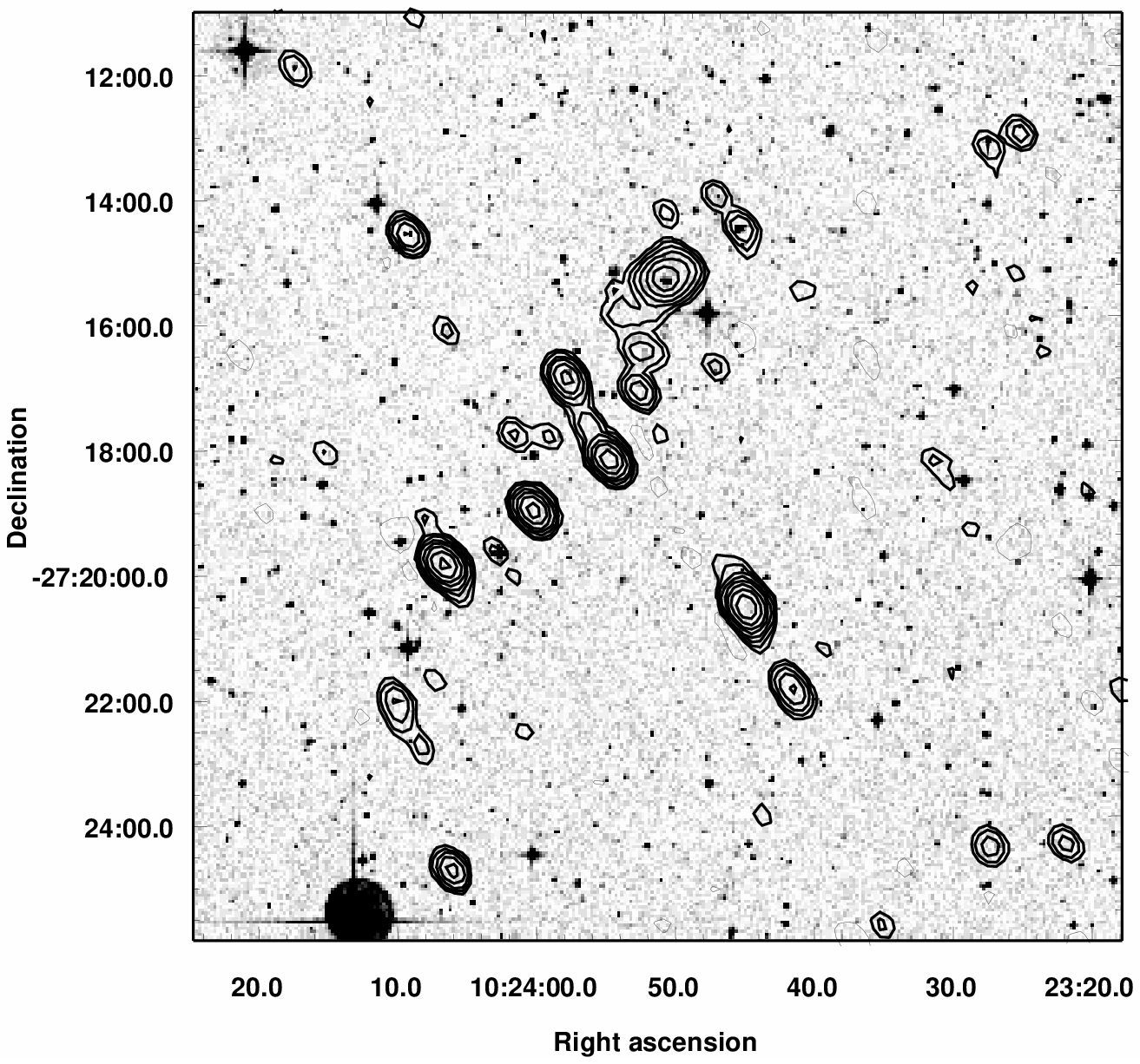}
\hspace{1truecm}\includegraphics[angle=0,width=7.6cm]{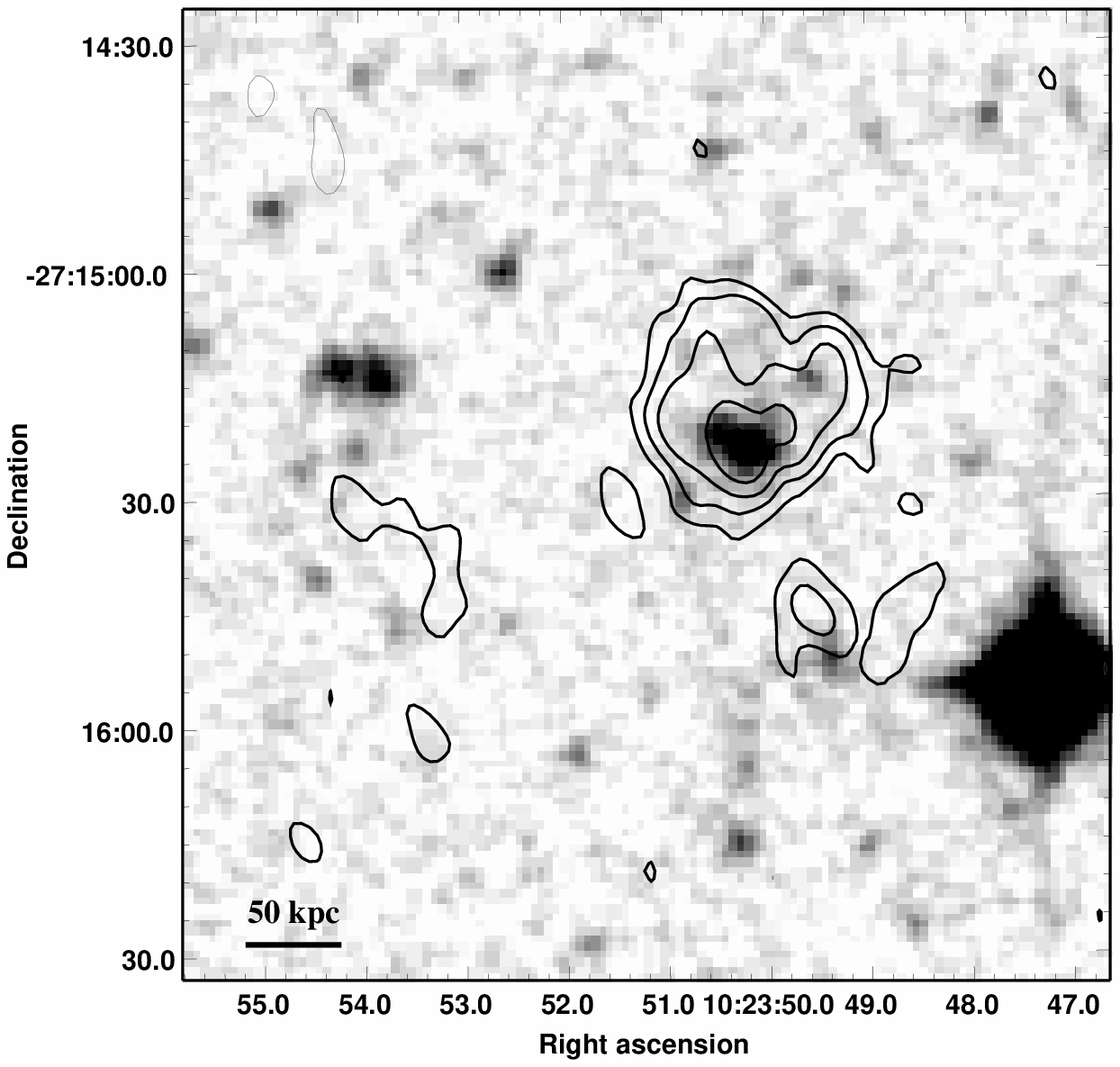}
\caption{Left -- GMRT 610 MHz radio contours for the cluster A\,3444
superposed on the POSS--2 optical plate.
The 1$\sigma$ level in the image is $\sim 50~\mu$Jy b$^{-1}$.  
Contours are 0.15$\times(\pm$1,2,4,8,16...)  mJy b$^{-1}$ (3$\sigma$). 
The HPWB is $23.2^{\prime\prime} \times 16.1^{\prime\prime}$, p.a.
$37.6^{\circ}$. Right -- High resolution zoom on the central cluster galaxy. 
The HPBW is $7.6^{\prime\prime} \times 4.9^{\prime\prime}$, 
p.a. $19^{\circ}$.
Contours are given starting from the 3$\sigma$ level: 
0.20$\times(\pm$1,2,4,8...).} 
\label{fig:a3444_opt}
\end{figure*}

Abell 3444 (RXCJ\,1023.8$-$2715, z=0.2542, 1$^{\prime\prime}$=3.924 kpc) was 
indicated as possible cooling core cluster
by L\'emonon (\cite{lemonon99}) and Matsumoto (\cite{matsumoto01}) on the basis
of the analysis of ASCA data, though at limited significance.
The X--ray ASCA image shows that the inner part of the cluster is elongated 
along the SE--NW direction.
\\
No radio information is reported in the literature. Unfortunately, due to
calibration problems, we could use only the USB of our dataset to image
this cluster. Our GMRT 610 MHz image
of the radio emission within the cluster virial radius is reported
in Fig. \ref{fig:a3444}, and shows that the radio emission is dominated by a
chain of individual sources, all with optical counterpart from the POSS--2.
The alignment of the chain of radio galaxies is in agreement with the inner  
elongation of the archive ASCA X--ray image. 
\\
A radio--optical overlay of the central part of the field is given
in Fig. \ref{fig:a3444_opt} (Left). The extended radio galaxy at 
the north--western end of the chain is associated with the dominant 
cluster galaxy (right panel in Fig. \ref{fig:a3444_opt}).
Its morphology is complex. Bent emission in the
shape of a wide angle tail is clear in the inner part of the source,
surrounded by extended emission. At least a couple of 
very faint objects are visible in the same region of the extended
radio emission, so it is unclear if we are dealing with extended
emission associated with the dominant cluster galaxy, or if this
feature is the result of a blend of individual sources.
Under the assumption that all the emission detected within the 
3$\sigma$ contour (left panel of Fig. \ref{fig:a3444_opt}) is associated 
with the dominant cluster galaxy, we measured a flux density  
S$_{\rm 610~MHz} = 16.5 \pm 0.8$ mJy,
which corresponds to logP$_{\rm 610~MHz}$(W/Hz) = 24.51. The largest 
angular size of the radio source is $\sim 40^{\prime\prime}$, hence
the linear size is $\sim$ 165 kpc.
\\
Both panels of Fig. \ref{fig:a3444_opt} suggest that emission on 
a larger scale may be present around the central radio source. Indeed
we measured a flux density of S$_{\rm 610~MHz} = 10.0\pm 0.8$ mJy on
an angular scale of $\sim 1.5^{\prime}$, i.e. $\sim$ 350 kpc.
\\
This situation is reminiscent of the class 
of core--halo sources, where extended emission surrounds a radio 
component obviously associated with a galaxy. 
Core--halo sources are usually located in cooling core clusters. 
Some well--known examples are 3C\,317 (Zhao, Sumi \& Burns \cite{zhao93}),
3C84 (B{\"o}hringer et al. \cite{boeringer93}), PKS 0745--191 (Baum \&
O'Dea \cite{baum91}). 
%
%

\subsection{Galaxy clusters without extended emission}\label{sec:noext}

For the remaining six clusters our 610 MHz GMRT observations did
not show any indication of possible extended emission at the noise
level of the final images.

\subsubsection{S\,780}

\begin{figure*}
\centering
\hspace{-0.5truecm}\includegraphics[angle=0,width=9.0cm,height=7.8cm]{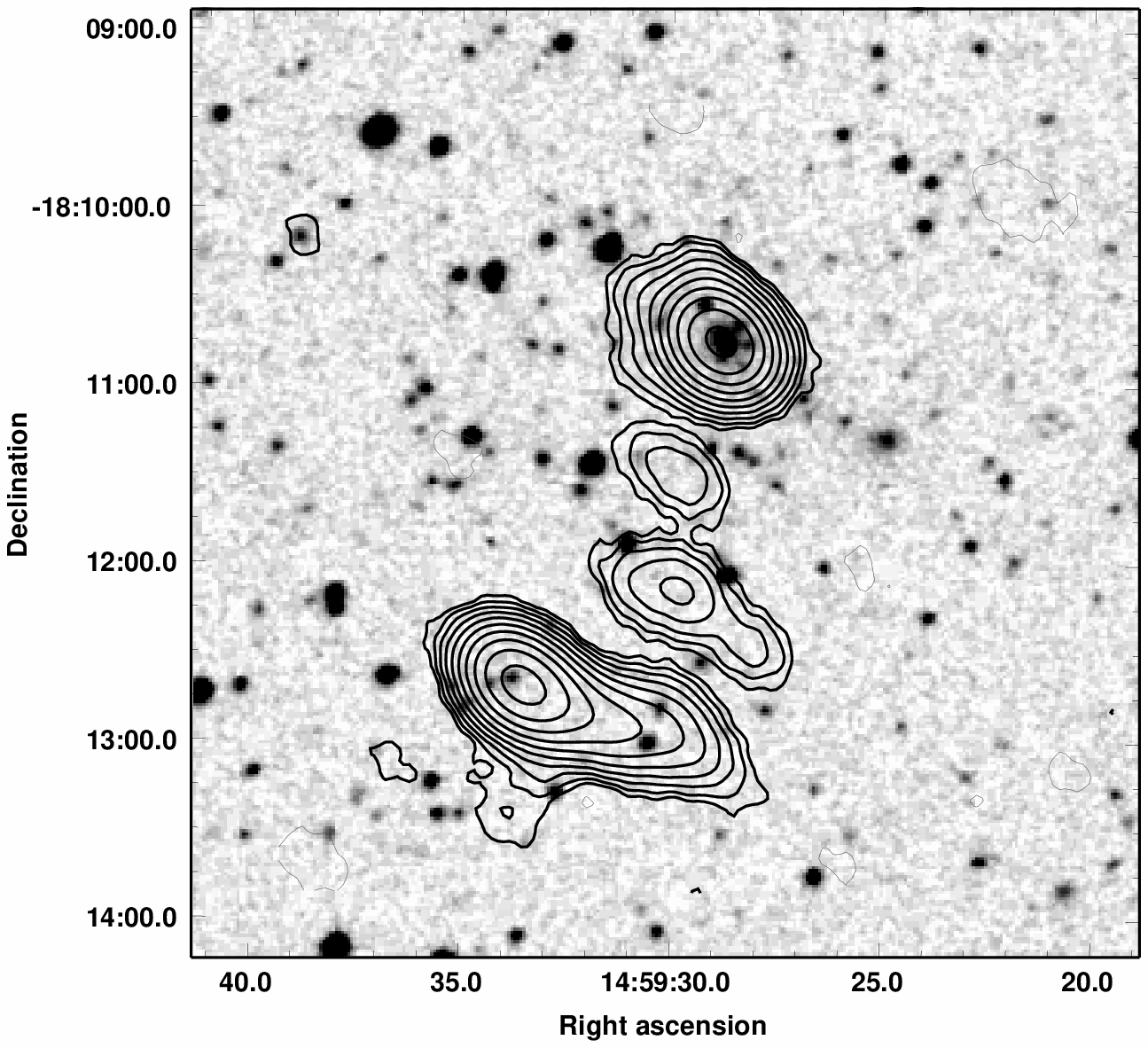}
\hspace{1truecm}\includegraphics[angle=0,width=8.3cm]{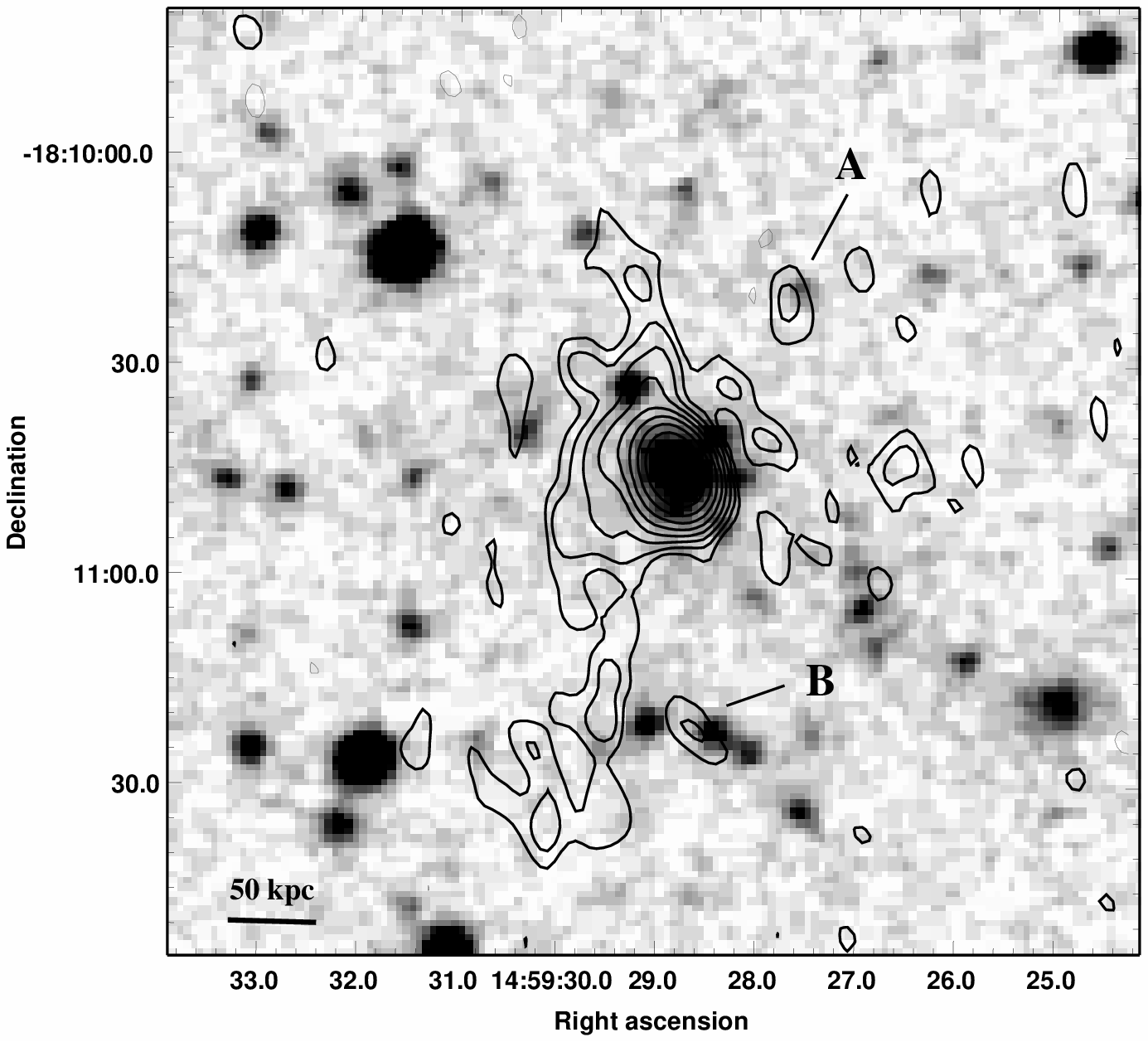}
\caption{Left -- GMRT 610 MHz radio contours for the cluster S\,0780
superposed on the POSS--2 optical plate.
The 1$\sigma$ level in the image is 65$~\mu$Jy b$^{-1}$. 
Contours are 0.20$\times(\pm$1,2,4,8,16...)  mJy b$^{-1}$. 
The HPWB is $20.9^{\prime\prime} \times 15.9^{\prime\prime}$, p.a.
$45.9^{\circ}$. Right -- High resolution zoom on the central cluster
galaxy. The HPWB is $7.5^{\prime\prime} \times 5.0^{\prime\prime}$, p.a.
$25^{\circ}$. Contours are 0.19$\times(\pm$1,2,4,8,16...) 
mJy b$^{-1}$ (first contous is 3$\sigma$).} 
\label{fig:s0780_opt}
\end{figure*}

S\,780 (RXCJ\,1459.4--1811, z=0.2357, 1$^{\prime\prime}$=3.952 kpc) is 
the most X--ray luminous and most massive cluster 
among those presented in this paper.
No information is available in the literature. 
Inspection of the ROSAT archive indicates that the X--ray emission  
is elongated in the E--W direction.
\\
Fig. \ref{fig:s0780} reports the 610 MHz contours of the S\,780 field
within the virial radius.
The radio emission from S\,780 is typical of a very active cluster,
with a number of cluster--type radio galaxies. 
Beyond the dominant central radio source, one head--tail radio
galaxy is clearly visible close to the cluster centre, one wide--angle tail 
is located at $\sim~6^{\prime}$ from the cluster centre (well within 
the virial radius) and one FRII radio galaxy (Fanaroff \& Riley \cite{fr74}) 
with distorted jets is located at  $\sim~8^{\prime}$ from the cluster centre 
(in the S--E direction). A few more radio
sources in the cluster field are optically identified. 
A visual inspection of the optical counterparts of all these radio sources 
suggests they have similar optical magnitudes.
\\
Radio--optical overlays are given in Fig. \ref{fig:s0780_opt}. The
left panel shows the central part of the cluster superposed on the 
POSS--2 optical frame, and the right panel is a high resolution zoom of
the central cluster galaxy. The radio galaxy shows a compact component
coincident with the nucleus of the associated galaxy, extended emission
in the eastern direction and a filament aligned South--East.
The total angular size is $\sim 50^{\prime\prime}$, corresponding to  
a largest linear size LLS $\sim$ 200 kpc. The flux density is 
S$_{\rm 610~MHz} = 135.9 \pm 6.8$ mJy, i.e. 
logP$_{\rm 610~MHz}$(W/Hz) = 25.32. Sources A and B highlighted in the right
panel of  Fig. \ref{fig:s0780_opt} were not included in the flux density
measurement. The flux density of the filament just South of the central 
radio source is S$_{\rm 610~MHz} = 3.1 \pm 0.2 mJy$.
No indication of residual flux density is present in the cluster centre.
\\
\\
%
%
%

\subsubsection{Abell 2697}
Very little information is available in the literature for
A\,2697 (RXCJ\,0003.1--0605, z=0.2320, 1$^{\prime\prime}$=3.698 kpc).
Archive X--ray ROSAT and ASCA images show that the hot intracluster
gas has a fairly regular distribution.
\\
For this cluster only one portion of the band (USB) was available. We 
imaged the cluster in a range of resolutions, 
reaching a 1$\sigma$ noise level of 
$\sim 80~\mu$Jy b$^{-1}$ in each image. 
The radio field is dominated by a head--tail galaxy.
Radio contours are reported in Fig. \ref{fig:a2697} in the Appendix. 
\\
No cluster--type extended feature is visible at the sensitivity 
level of the images, and no significant flux density from positive 
residuals was found by integrating over a wide region of the 
cluster centre.
\\

\subsubsection{Abell 141}

The X--ray emission of A\,141 (RXCJ\,0105.5--2439, z=0.2300,
1$^{\prime\prime}$=3.674 kpc) is bimodal. The archive X--ray ROSAT 
images show that the North--South elongation of the ASCA image is the 
result of two components, the northern one being the brightest and 
largest. The same orientation was found also in the
distribution of the cluster galaxies by Dahle et al. (2002), who concluded
that the overall optical analysis is suggestive of recent
merger activity. Those authors reported also evidence of weak lensing.
\\
High resolution radio imaging aimed at the detection of cluster radio
galaxies was carried out with the VLA--A at 1.4 GHz by Rizza et al.
(\cite{rizza03}).
\\
For this cluster only one portion of the observing band (USB) was 
available. 
Our GMRT observations of A\,141 revealed neither the presence of diffuse 
emission at the level of $\sim 100~\mu$Jy b$^{-1}$, nor unusually high residuals. 
Radio contours are reported in Fig. \ref{fig:a141} in the Appendix.

\subsubsection{Abell 3088}

Very little information is available in the literature for 
A3088 (RXCJ\,0307.0$-$2840, z=0.2537, 1$^{\prime\prime}$=3.952 kpc). 
It is a richness 2 galaxy cluster with a
very regular and symmetric X--ray morphology. On the basis of XMM--Newton
observations, Zhang et al. (\cite{zhang06}) reported a gas temperature 
kT=6.4$\pm$0.3 keV and classified it as a ``single dynamical state'' cluster
with a cooling core (Finoguenov, B{\"o}hringer \& Zhang \cite{finogue05}).
\\
Our GMRT 610 MHz observations show that the field has only a few radio sources,
with a lack of positive residual flux density in the central cluster region 
and no hints  of diffuse emission from the cluster at the detection level 
of our images, i.e  1$\sigma~\sim~65~\mu$Jy b$^{-1}$.
Contours of the radio emission within the cluster virial radius  
are shown in Fig. \ref{fig:a3088}.

\subsubsection{Abell 2537}

Little information is available in the literature for 
A\,2537 (RXCJ\,2308.2--0211, z=0.2966, 1$^{\prime\prime}$=4.419 kpc).
Archive HST observations show the presence of several red and blue arcs,
and Dahle et al. (\cite{dahle02}) report evidence of weak lensing.
The cluster was observed in the X--ray band by XMM--Newton 
and was classified as ``single dynamical state'', with gas temperature 
kT=7.9$\pm$0.7 KeV (Zhang et al. \cite{zhang06}). 
A secondary X--ray peak is present at $\sim~7^{\prime}$ from the 
cluster gas concentration.
\\
For this cluster only the USB provided useful data.
The 610 MHz radio emission from the cluster, shown in Fig. \ref{fig:a2537}
is dominated by a tailed radio galaxy located at the cluster centre.
Very few other radio sources are detected above the 5$\sigma$ level of
the image. No hint of extended emission is present in the field
at the level of $\sim 60~ \mu$Jy b$^{-1}$ (1$\sigma$), and no 
high positive flux density residuals were detected over the central 
cluster region.
\\

\subsubsection{Abell 2631}

Little information is available in the literature for the rich cluster 
A\,2631 (RXCJ\,2337.6+0016, R=3, z=0.2779, 1$^{\prime\prime}$=4.221 kpc).
Archive ROSAT X--ray images are available for A\,2631, which show a
complex morpholgy. Based on XMM--Newton observations Zhang et al.
(\cite{zhang06}) classified it as ``offset centre'', with varying 
isophote centroids on different angular scales, and reported 
a gas temperature kT=9.6$\pm$0.3 KeV. Finoguenov et al. (\cite{finogue05})
interpret the XMM properties of this cluster in terms of a late stage
of a core disruption.
The cluster was observed with the VLA--A at 1.4 GHz
(Rizza et al. \cite{rizza03}).
\\
Our GMRT observations of this cluster were spread over two days,
however on both days only one portion of the band (USB) was available.
The 610 MHz radio emission of A\,2631 within the virial radius is 
shown in Fig. \ref{fig:a2631}. It is dominated by a central
tailed radio galaxy and all the remaining sources above the 5$\sigma$
level are located South of the cluster centre. No signs of extended emission
are present in the field at the rms level of $\sim 50~ \mu$Jy b$^{-1}$ 
(1$\sigma$), and no positive residuals were found by integrating over the
central region of the cluster.
\\

\section{Discussion and conclusions}\label{sec:discussion}

Our 610 MHz GMRT radio halo survey has been designed in
order to statistically investigate the connection 
between cluster merger phenomena and the presence of cluster--scale
radio emission. In particular, our main goal is to derive the fraction 
of massive clusters (i.e. $M \ge 10^{15} M_{\odot}$) with 
giant radio halos in the redshift range 0.2$<$z$<$0.4, in order to constrain 
the expectations made by CBS04 and CB05 in the framework of the 
particle re--acceleration model.
The total cluster sample consists of two sub--samples of massive clusters
extracted from the REFLEX and extended BCS catalogues, and includes a total 
of 50 clusters.
\\
The cluster sample presented here (see Table 1) includes 27 REFLEX
clusters, eleven of which were observed in a first run of GMRT observations
carried out in January 2005. If we consider the literature data, information
is now available for 15 of the 27 objects.  
The most relevant results we obtained, as well as the status of the observations for 
the remaining clusters in the sample are summarized below.

\begin{itemize}

\item[{\it (a)}] Two new giant radio halos were found,
i.e. A\,209 (also reported in Giovannini et al. \cite{gg06} while
this paper was in preparation), and RXCJ\,2003.5--2323, discovered with the 
present 610 MHz GMRT observations.

\item[{\it (b)}] A radio halo (LLS$\sim$460 kpc) was found in 
RXCJ\,1314.4--2515.

\item[{\it (c)}] Two relics were found in the cluster RXCJ\,1314.4--2515,
and one  in A\,521 (Giacintucci et al. \cite{giacintucci06}).
These three relics are impressive structures. 
Their largest linear size is of the order of the Mpc, which suggests
that particle acceleration, most likely related to the hierarchical
formation of clusters and accretion processes, might be required to
account for their formation (e.g. Ensslin \& Br\"uggen \cite{eb02}).
The relic in A\,521 has already been
studied in detail. Here we just wish to mention that RXCJ\,1314.4--2515 
is the third galaxy cluster known to date hosting two relics, 
after A\,3667 (Roettiger, Burns \& Stone \cite{roetti99};
Johnston--Hollitt et al. \cite{jh02}) and A\,3376 (Bagchi et al. \cite{bagchi06}). 
Furthermore, it is unique in hosting two relic sources and one radio halo, 
and hence a challenge for our understanding of the connection between radio 
halos, relics and the physics of cluster mergers.

\item[{\it (d)}] Extended emission on smaller scale (of the order of $\sim 350$ kpc)
was detected around the dominant galaxy 
in A\,3444, whose radio morphology and monocromatic power are similar to those 
of core--halo radio galaxies found at the centre of cooling core clusters.

\item[{\it (e)}] Three clusters in the sample host  well--known giant 
radio halos, i.e. A2744 (Govoni et al. \cite{govoni01}), A1300 
(Reid et al. \cite{reid99}) and A2163 (Herbig \& Birkinshaw \cite{herbig94}, 
Feretti et al. \cite{feretti01}).

\item[{\it (f)}] No extended emission of any kind was detected at the level
of 50 -- 100 $\mu$Jy b$^{-1}$ in six of the 11 clusters observed by us and
presented here.

\item[{\it (g)}] The cluster RXCJ\,0437.1+0043 is known not to host extended emission,
based on low resolution 1.4 GHz VLA observations (Feretti et al. 
\cite{feretti05}).

\item[{\it (h)}] Five clusters were observed by us in October 2005, two more will be 
observed in August 2006, and they will be presented in a future paper 
(see Sect. \ref{sec:obs}).

\item[{\it (h)}] The remaining 5 clusters are being observed by other authors
(GMRT Cluster Key Project, P.I. Kulkarni) and no literature information 
is available thus far.

\end{itemize}

%
\begin{figure}
\centering
\includegraphics[angle=0,width=8.5cm]{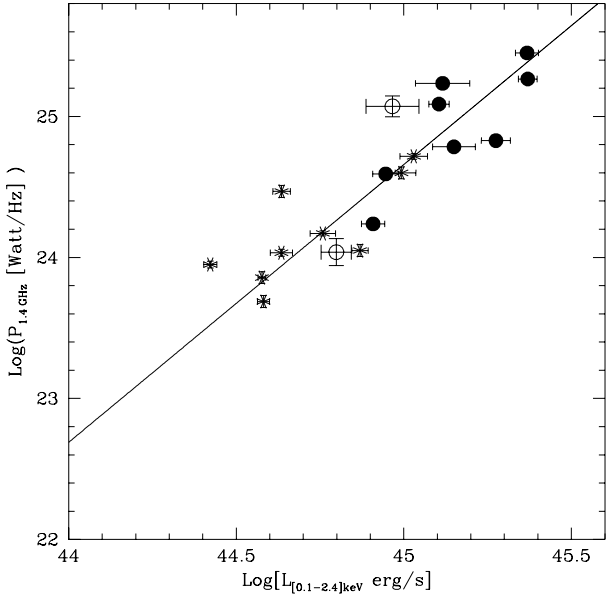}
\caption{LogL$_{\rm X}$--LogP$_{\rm 1.4~GHz}$ plot for the clusters
with detected giant radio halos. Stars represent the literature clusters 
at z$<$0.2 and filled circles the literature clusters at z$>0.2$.   
Open circles show the location of A\,209 (lower left) 
and RXCJ\,2003.5--2323 (upper right).}
\label{fig:LxLr}
\end{figure}

In Fig. \ref{fig:LxLr} 
we show the location of the giant radio halos in A\,209 and
RXCJ\,2003.5--2323 in the log~L$_{\rm X}$ -- log(P$_{\rm 1.4~GHz}$) plane,
where all the previously known clusters with giant radio halos are also 
reported (see CBS06 and references therein for the literature data). 
The radio power at
1.4 GHz for these two clusters was obtained scaling the measured flux density
at 610 MHz with a spectral index $\alpha_{\rm 610~MHz}^{\rm 1.4~GHz} = 
1.2 \pm 0.2$ (the uncertainty assumed here dominates over the 610 MHz
flux density error). Clusters at z$<$0.2 and those at z$>$0.2 are shown with 
different symbols. Despite some overlap, the most powerful radio halos are 
hosted in the most X--ray luminous clusters, which are also the most distant.
The location of A\,209 and RXCJ2003.5--2323 on the plot is in good agreement
with the distribution of all giant radio halos known in the literature. 
\\
\\
An important piece of information would be the knowledge of the 
merging stage of the clusters in the sample, since cluster merger is a major
ingredient in the re--acceleration model.
The literature information on the clusters presented here is not
homogenous, and it is not possible to make conclusive statements
on the connection between merging/non--merging signatures and the
presence/absence of radio halos. A\,209 is known to be undergoing merging 
events, but no information is available for RXCJ\,2003.5--2323, except
for the elongated X--ray emission imaged by ROSAT. The three
radio halo clusters known from the literature are all reported to
be dynamically active  
(see for instance Zhang et al. \cite{zhang06} and Finoguenov et 
al. \cite{finogue05}). Signature 
of cluster merger is present in the optical and X--ray bands 
for A\,521 (Giacintucci et al. \cite{giacintucci06} and references therein)
and RXCJ\,1314.4--2515, which host extended radio
emission in the form of radio halo and relics. Elongated or more
complex X--ray isophotes are visible in S\,780, A\,141, A\,2631 
and in RXCJ\,0437.1+0043, which lack cluster scale 
radio emission.
The remaining two clusters without extended emission are considered
``relaxed'' on the basis of the X--ray emission
\\
To summarize, the optical and X--ray information for the sample of 
clusters presented here is not inconsistent with the findings that clusters
with radio halos are characterised by signatures of merging processes. 
On the other hand, clusters without extended radio emission may or may not
show dynamical activity at some level.
This crucial issue will be further investigated in future works.
\\
\\
In the framework of the canonical particle re--acceleration model
giant radio halos are believed to be essentially
an on/off phenomenon, triggered by dissipation via collisionless
damping of turbulence injected during cluster mergers.
\\
The physics of collisionless turbulence and of particle acceleration is 
still poorly understood and many hidden ingredients could be of relevance
in computing the efficiency of the particle acceleration processes
in the ICM.
On the other hand, from simple energetic arguments, 
it is clear that the possibility to develop a giant radio halo 
is related to the efficiency of turbulence injection and to the   
possibility to generate large enough ($\geq$Mpc sized) turbulent cluster 
regions. In this respect, the calculations in CBS04, CB05 and more recently
in CBS06 show that major cluster mergers (i.e. with  mass ratio of the 
order $\leq$ 5:1) between massive clusters (M$\geq 10^{15}$M$_{\odot}$) may 
provide the necessary ingredients to develop giant radio halos.
During these mergers a fraction of up to $\sim 10\%$ of the cluster
thermal energy is believed to be injected in a $\sim {\rm Mpc}^3$ region.
However, from a theoretical point of view, it is hard to predict if a 
particular merging 
clusters may host a giant radio halo, since this depends 
on a number of parameters which cannot be easily estimated.
For instance, in order to have enough time for the turbulence injected 
on large scales to cascade down to collisionless scales, it is necessary 
that seed relativistic particles (to be reaccelerated) are present
in the turbulent ICM, and that the magnetic field in the ICM is strong 
enough to allow $\sim$GeV electrons to emit synchrotron radiation at the 
observed frequency.
\\
The statistical approach developed in CBS04, CB05 and CBS06 allows 
a more reliable estimate of the fraction of clusters hosting a giant
radio halo.
Without going into the calculation details, the most
relevant result in the light of those papers is that the  
fraction of galaxy clusters with mass  
$\sim 2-3.5\times 10^{15} M_{\odot}$ and redshift $z = 0.2 \div 0.4$ 
expected to host a giant radio halo is in the range 
$\sim 10-35 \%$.
In addition, CBS06 showed that the cluster magnetic field plays
an important role, and that this fraction depends 
the scaling law between the magnetic field and cluster mass. 
\\
Radio information is now available for 15 of the 27 
clusters considered in this paper, and 5 of them possess a giant radio 
halo. However, at this stage of our work the statistics are still poor, 
and no firm comparison with theoretical expectations can be reached.
For this reason our analysis, in the light of 
the predictions made in CB05 and CBS06, will be carried out as soon
as the information on the whole selected sample (REFLEX and BCS) is
completed (Venturi et al. in prep.; Cassano et al in prep.).
%
\\ 
\\
{\it Acknowledgements.}
We thank the staff of the GMRT for their help during the observations.
GMRT is run by the National Centre for
Radio Astrophysics of the Tata Institute of Fundamental Research.
T.V. and S.G. acknowledge partial support from the Italian Ministry
of Foreign Affairs. G.B., R.C. and G.S. acknowledge partial support 
from MIUR grants PRIN2004 and PRIN2005. 

\appendix
\section{Radio Images}

In this appendix we report the 610 MHz radio contours 
of all the observed clusters in the present project. The
images cover the region within the cluster virial radius. 
The resolution is 15.0$^{\prime \prime} \times 12.0^{\prime \prime}$
in all the images, except for A209, RXCJ\,1314.4--2515 and 
RXCJ\,2003.5--2323 (see figure caption).

%
%
\begin{figure*}
\centering
\includegraphics[angle=0,width=12.5cm]{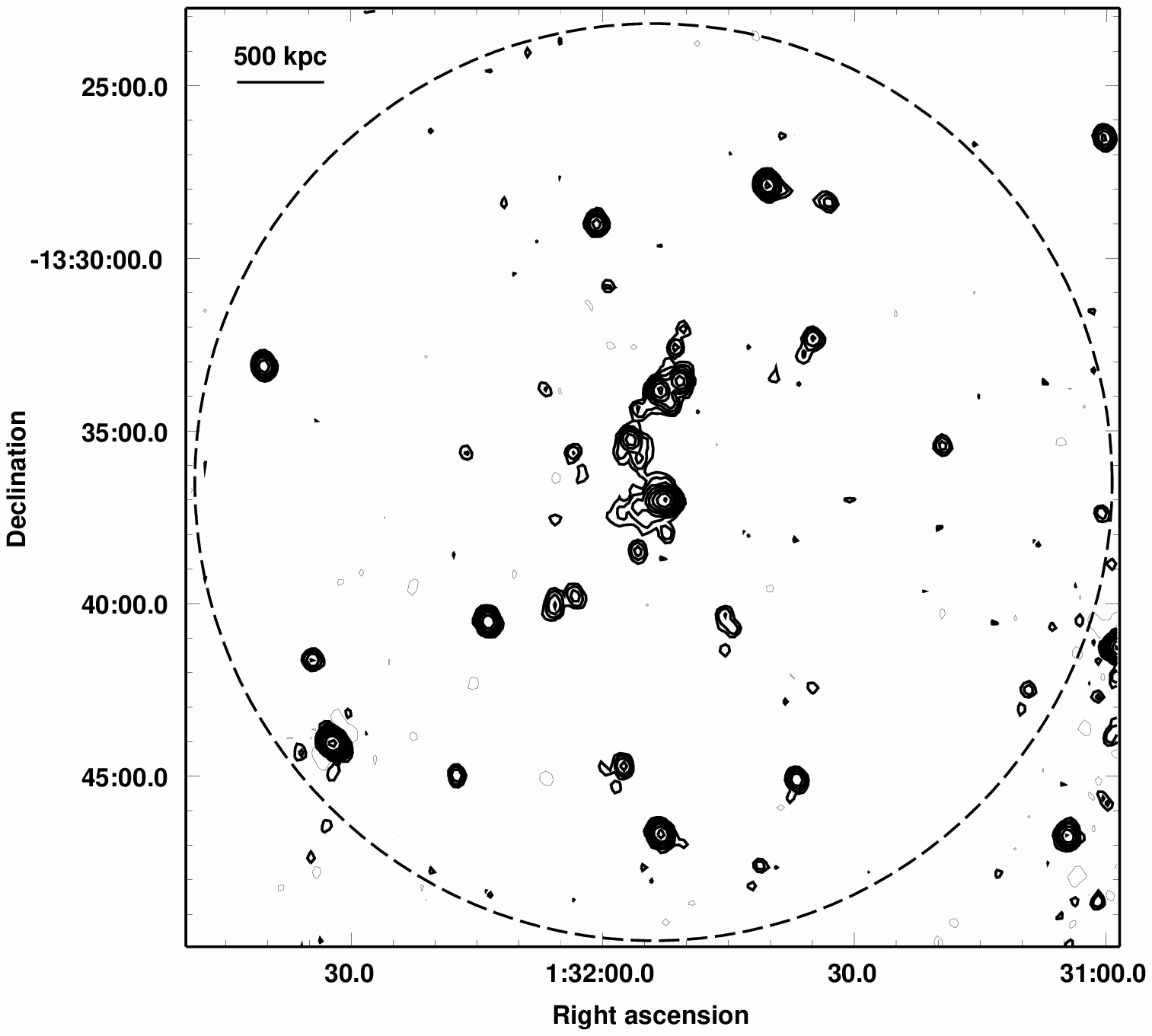}
\caption{GMRT 610 MHz radio contours for the cluster A\,209.
The 1$\sigma$ level in the image is 50$\mu$Jy b$^{-1}$.  
Contours are 0.20$\times(\pm$1,2,4,8,16...) mJy b$^{-1}$. The 
HPWB is $21.0^{\prime\prime} \times 18.0^{\prime\prime}$, p.a.
$13^{\circ}$. The radius of the dashed circle is the 
virial radius, corresponding to 13.28$^{\prime}$ for this cluster.}
\label{fig:a209_lr}
\end{figure*}
%

%
%
\begin{figure*}
\centering
\includegraphics[angle=0,width=13cm]{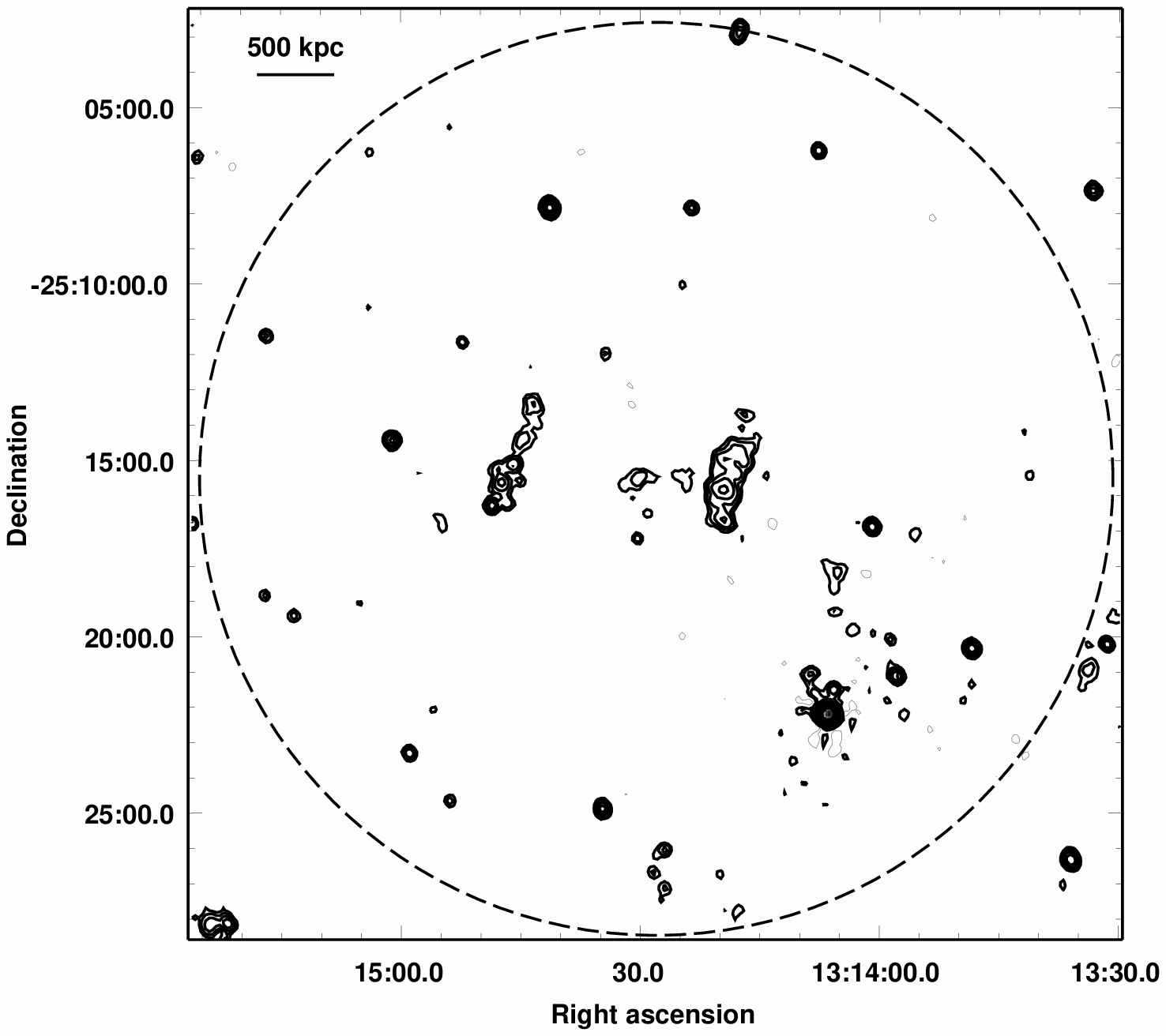}
\caption{GMRT 610 MHz radio contours for the cluster RXCJ\,1314.4--2515.
The 1$\sigma$ level in the image is 60$\mu$Jy b$^{-1}$.  
Contours are 0.3$\times(\pm$1,2,4,8,16...) mJy b$^{-1}$. The 
HPWB is $15.0^{\prime\prime} \times 13.0^{\prime\prime}$, p.a.
$15^{\circ}$. The radius of the dashed circle is the 
virial radius, corresponding to 12.94$^{\prime}$ for this cluster.}
\label{fig:rxcj1314_lr}
\end{figure*}
\begin{figure*}
\centering
\includegraphics[angle=0,width=12.5cm]{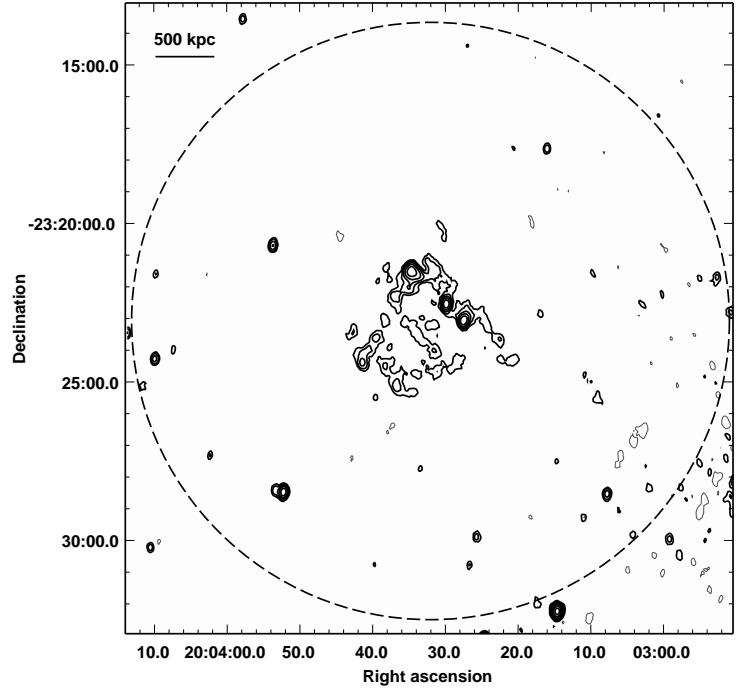}
\caption{GMRT 610 MHz radio contours for the cluster RXCJ\,2003.5--2323.
The 1$\sigma$ level in the image is 60$\mu$Jy b$^{-1}$.  
Contours are 0.3$\times(\pm$1,2,4,8,16...) mJy b$^{-1}$. The 
HPWB is $15.5^{\prime\prime} \times 10.3^{\prime\prime}$, p.a.
$-6^{\circ}$. The radius of the dashed circle is the 
virial radius, corresponding to 9.91$^{\prime}$ for this cluster.}
\label{fig:rxcj2003}
\end{figure*}
\begin{figure*}
\centering
\includegraphics[angle=0,width=12.3cm]{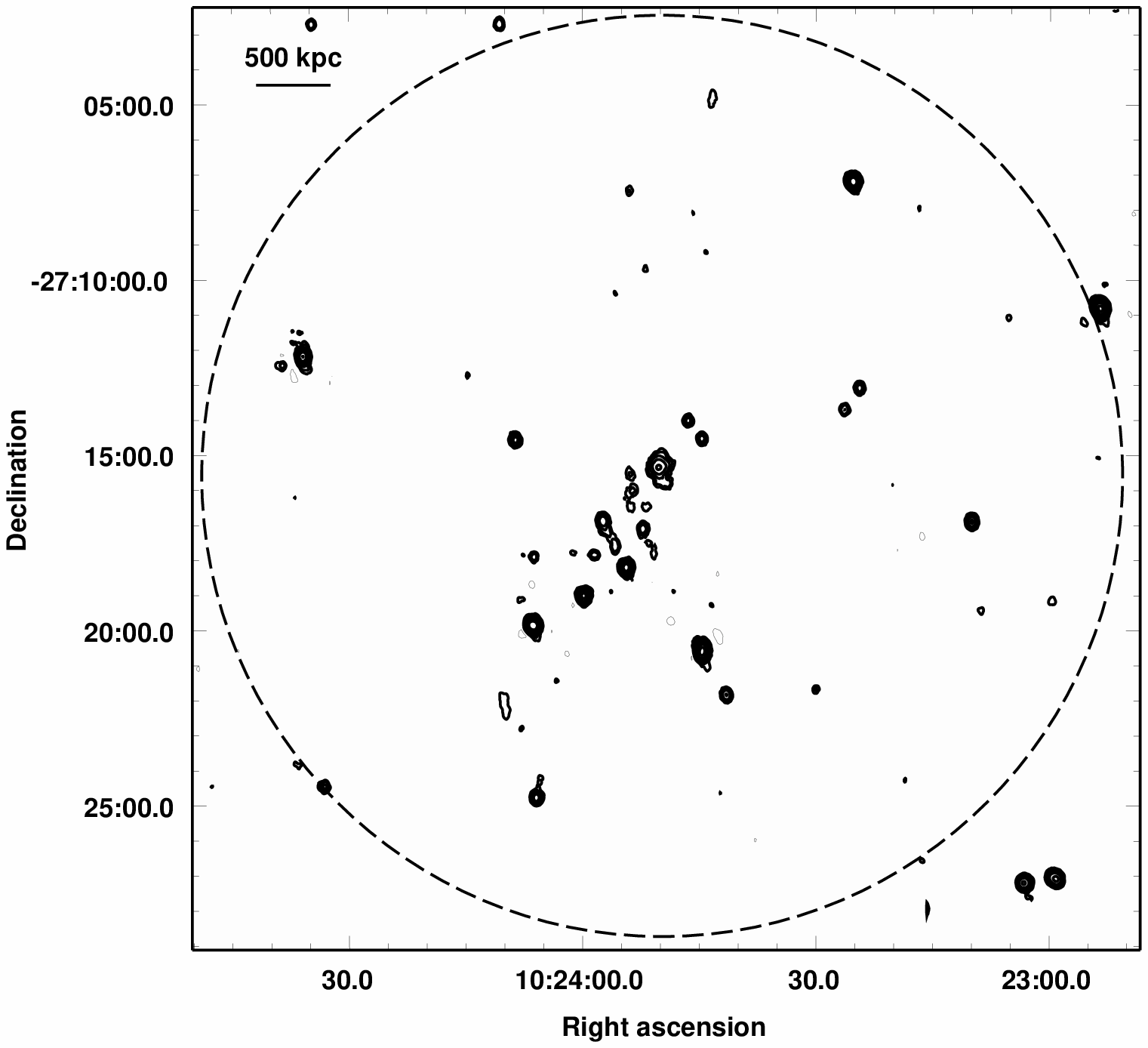}
\caption{GMRT 610 MHz radio contours for the cluster A\,3444.
The 1$\sigma$ level in the image is 60$\mu$Jy b$^{-1}$. 
Contours are 0.4$\times(\pm$1,2,4,8,16...) mJy b$^{-1}$. The 
HPWB is $15^{\prime\prime} \times 12^{\prime\prime}$, p.a.
$0^{\circ}$. The radius of the dashed circle is the 
virial radius, corresponding to 13.14$^{\prime}$ for this cluster.}
\label{fig:a3444}
\end{figure*}

\begin{figure*}
\centering
\includegraphics[angle=0,width=12.5cm]{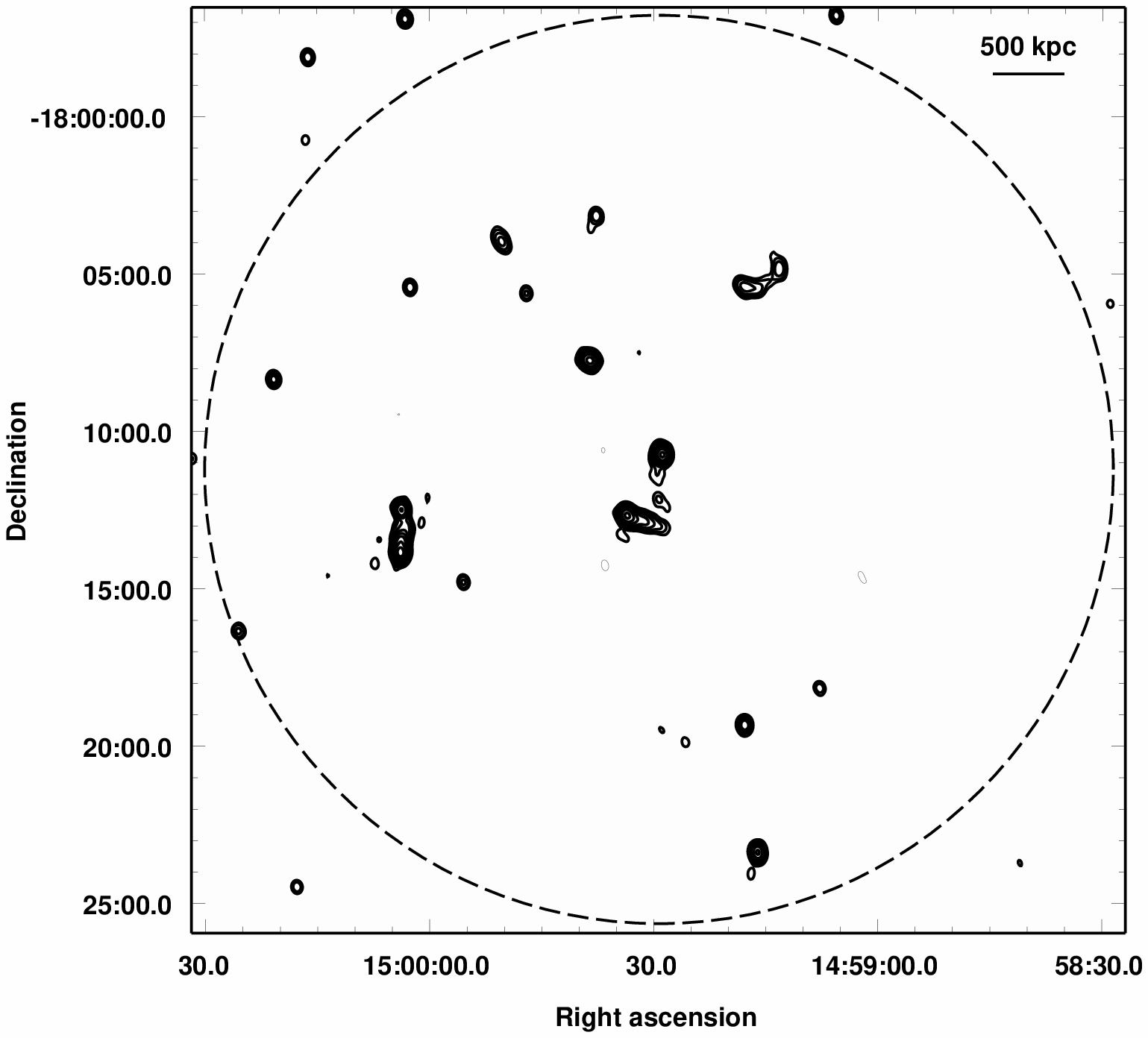}
\caption{GMRT 610 MHz radio contours for the cluster S\,780.
The 1$\sigma$ level in the image is 65$\mu$Jy b$^{-1}$.  
Contours are 0.3$\times(\pm$1,2,4,8,16...) mJy b$^{-1}$. The 
HPWB is $15^{\prime\prime} \times 12^{\prime\prime}$, p.a.
$0^{\circ}$. The radius of the inner circle is the 
virial radius, corresponding to 14.43$^{\prime}$ for this cluster.}
\label{fig:s0780}
\end{figure*}

\begin{figure*}
\centering
\includegraphics[angle=0,width=12.5cm]{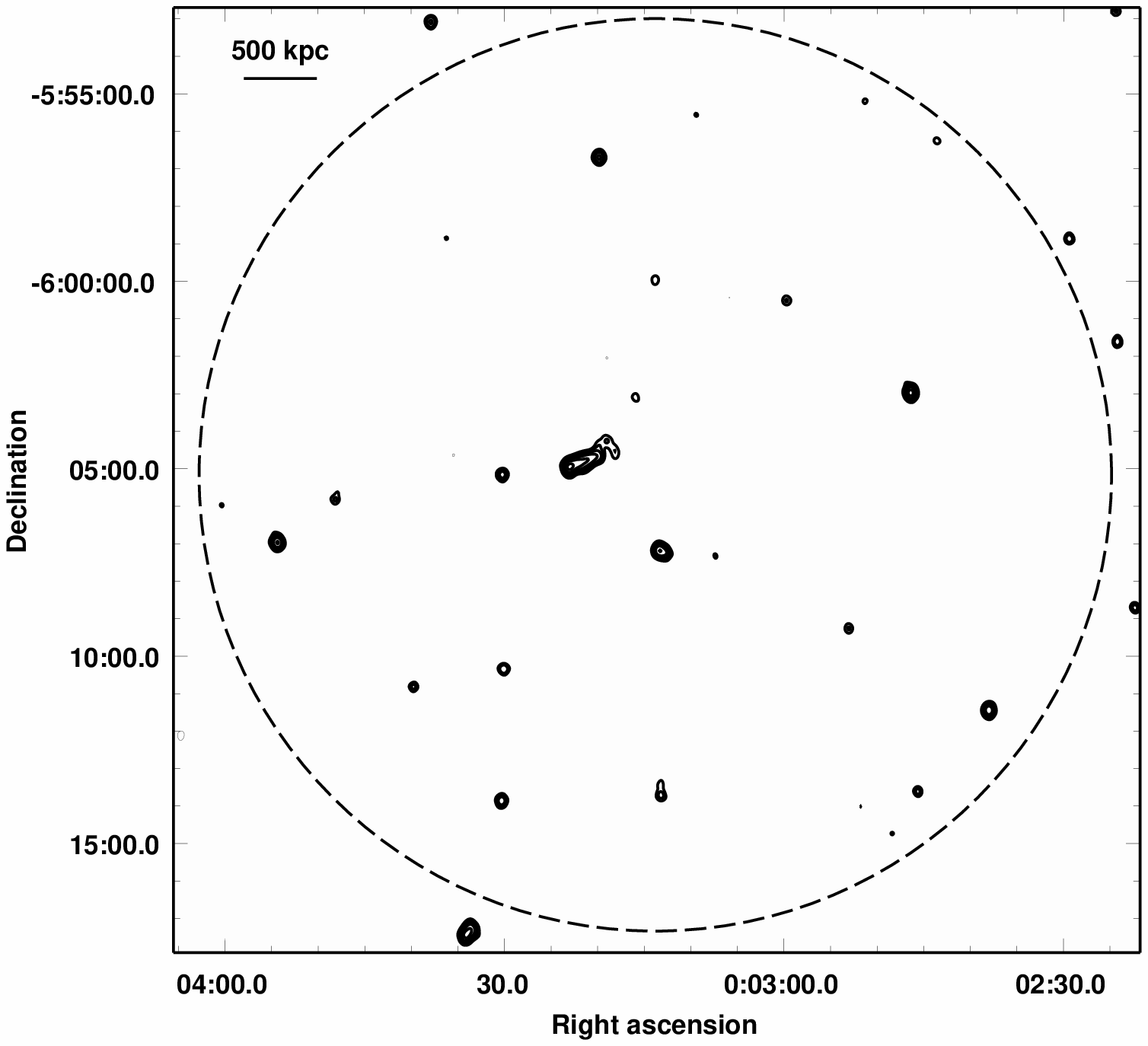}
\caption{GMRT 610 MHz radio contours for the cluster A\,2697.
The 1$\sigma$ level in the image is 80$\mu$Jy b$^{-1}$. 
Contours are 0.4$\times(\pm$1,2,4,8,16...) mJy b$^{-1}$. The 
HPWB is $15^{\prime\prime} \times 12^{\prime\prime}$, p.a.
$0^{\circ}$. The radius of the dashed circle is the 
virial radius, corresponding to 12.17$^{\prime}$ for this cluster.}
\label{fig:a2697}
\end{figure*}

\begin{figure*}
\centering
\includegraphics[angle=0,width=12.5cm]{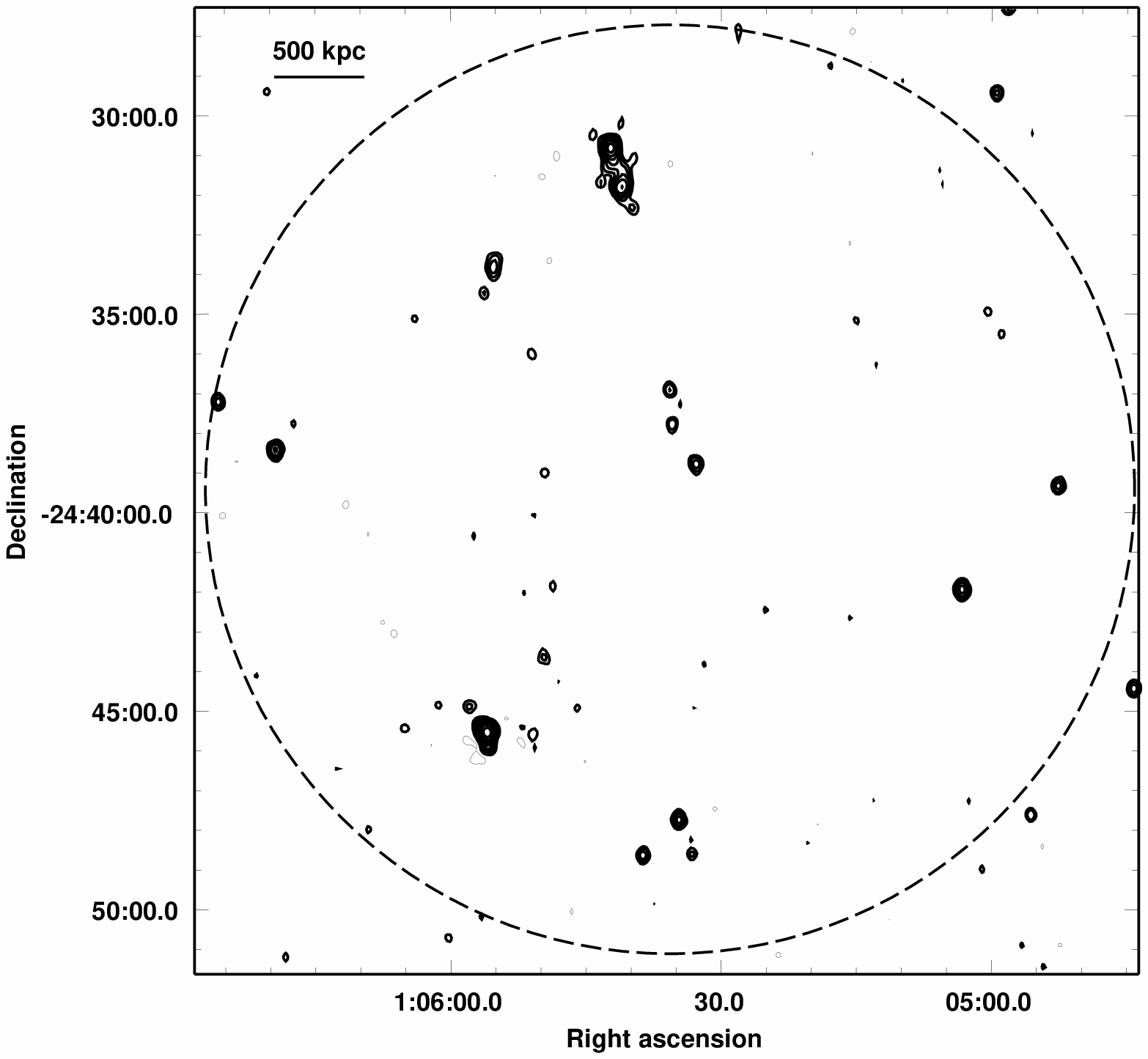}
\caption{GMRT 610 MHz radio contours for the cluster A\,141.
The 1$\sigma$ level in the image is 90$\mu$Jy b$^{-1}$. 
Contours are 0.45$\times(\pm$1,2,4,8,16...) mJy b$^{-1}$. The 
HPWB is $15^{\prime\prime} \times 12^{\prime\prime}$, p.a.
$0^{\circ}$. The radius of the dashed circle is the 
virial radius, corresponding to 11.79$^{\prime}$ for this cluster.}
\label{fig:a141}
\end{figure*}

\begin{figure*}
\centering
\includegraphics[angle=0,width=12.5cm]{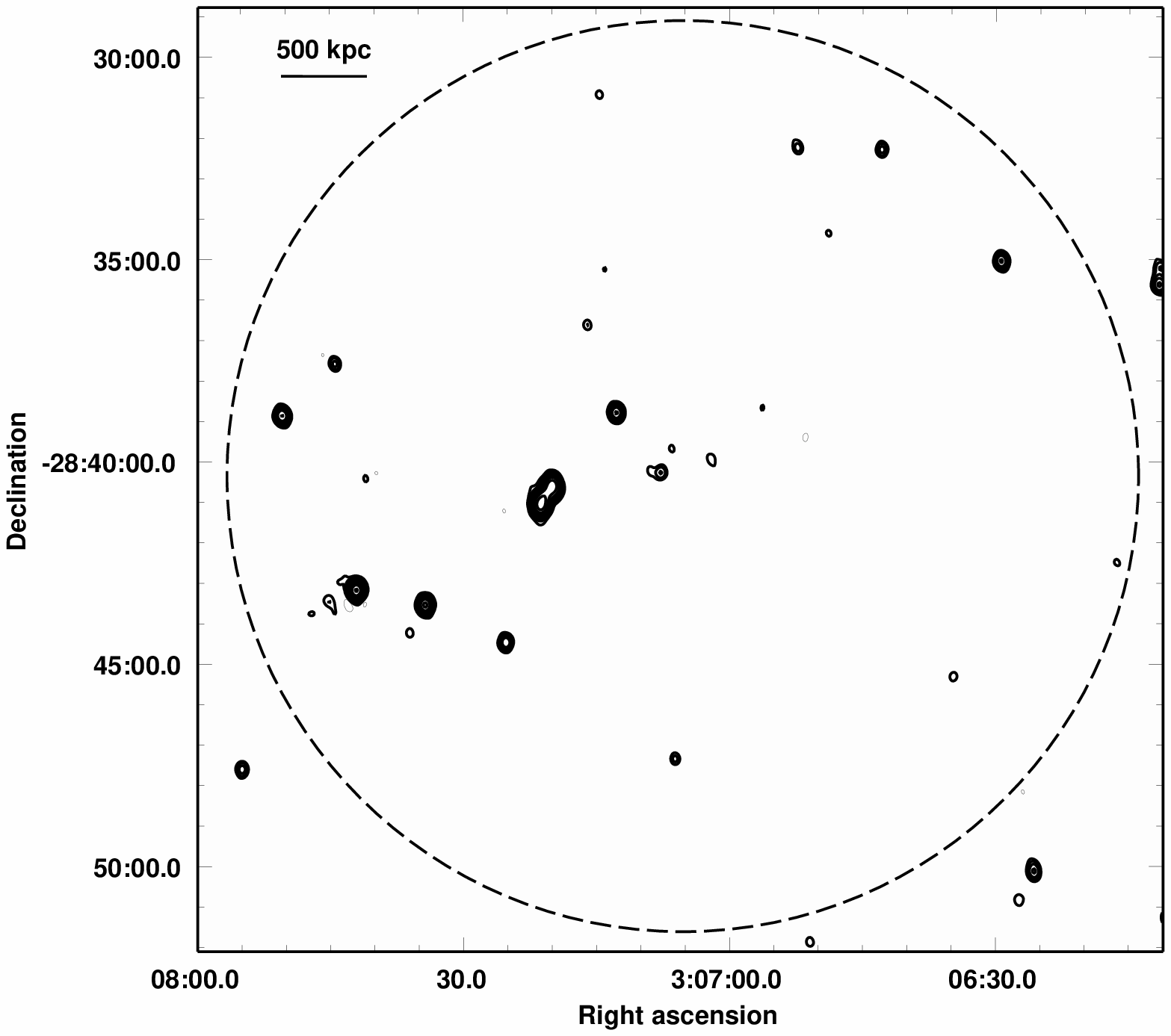}
\caption{GMRT 610 MHz radio contours for the cluster A\,3088.
The 1$\sigma$ level in the image is 65$\mu$Jy b$^{-1}$. 
Contours are 0.325$\times(\pm$1,2,4,8,16...) mJy b$^{-1}$. The 
HPWB is $15^{\prime\prime} \times 12^{\prime\prime}$, p.a.
$0^{\circ}$. The radius of the dashed circle is the 
virial radius, corresponding to 11.36$^{\prime}$ for this cluster.}
\label{fig:a3088}
\end{figure*}

\begin{figure*}
\centering
\includegraphics[angle=0,width=12.5cm]{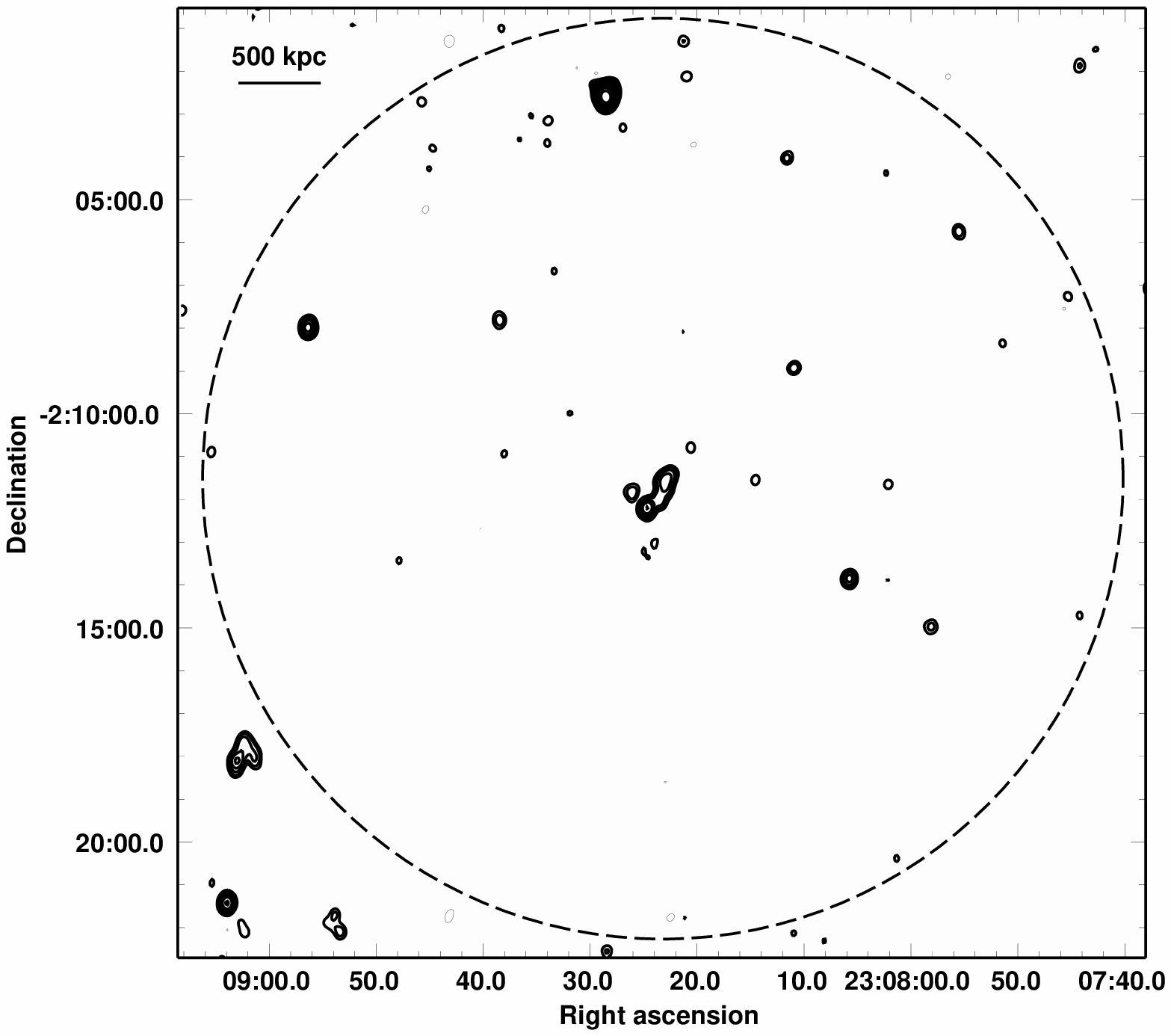}
\caption{GMRT 610 MHz radio contours for the cluster A\,2537.
The 1$\sigma$ level in the image is 80$\mu$Jy b$^{-1}$. 
Contours are 0.4$\times(\pm$1,2,4,8,16...) mJy b$^{-1}$. The 
HPWB is $15^{\prime\prime} \times 12^{\prime\prime}$, p.a.
$0^{\circ}$. The radius of the dashed circle is the 
virial radius, corresponding to 10.75$^{\prime}$ for this cluster.}
\label{fig:a2537}
\end{figure*}

\begin{figure*}
\centering
\includegraphics[angle=0,width=12.5cm]{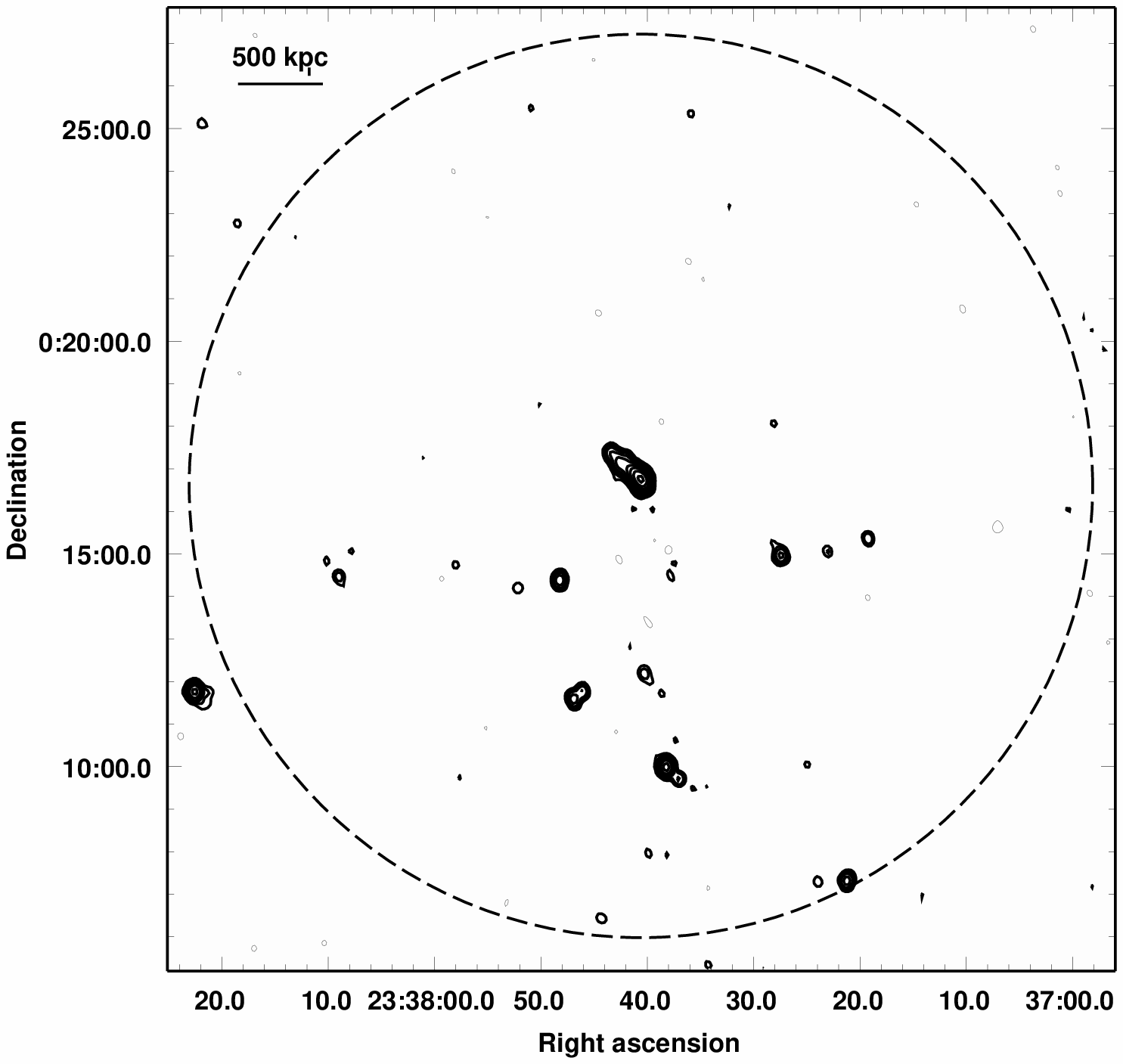}
\caption{GMRT 610 MHz radio contours for the cluster A\,2631.
The 1$\sigma$ level in the image is 80$\mu$Jy b$^{-1}$. 
Contours are 0.4$\times(\pm$1,2,4,8,16...) mJy b$^{-1}$. The 
HPWB is $15^{\prime\prime} \times 12^{\prime\prime}$, p.a.
$0^{\circ}$. The radius of the dashed circle is the 
virial radius, corresponding to 10.62$^{\prime}$ for this cluster.}
\label{fig:a2631}
\end{figure*}

\end{document}